\documentclass[sigconf,nonacm]{acmart}

\setcopyright{none}
\renewcommand\footnotetextcopyrightpermission[1]{}
\usepackage{booktabs}
\usepackage{graphicx}
\usepackage{amsmath}

\usepackage{amssymb}
\usepackage{xcolor}
\usepackage{algorithm}
\usepackage{algpseudocode}
\usepackage{xspace}
\usepackage{float}
\newcommand{\rgd}{\textsc{RigidShift}\xspace}

\begin{document}

\title{Rigid-Covert GNSS Spoofing of UAV Swarms: A Structural Blind Spot,
Its Detection Limit, and Absolute-Anchor Defenses}

\author{Minseok Park}
\affiliation{%
  \institution{Jeonbuk National University}
  \department{Dept. of Electronic Engineering}
  \city{Jeonju}
  \country{Republic of Korea}}
\email{yp1278kr@jbnu.ac.kr}

\author{Joon Soo Yoo}
\authornote{Corresponding author.}
\affiliation{%
  \institution{Jeonbuk National University}
  \department{Dept. of Advanced Defense Technology and Industry}
  \city{Jeonju}
  \country{Republic of Korea}}
\email{joonsooyoo@jbnu.ac.kr}
\renewcommand{\shortauthors}{Park and Yoo}

\begin{abstract}
Cooperative UAV-swarm defenses commonly cross-check GNSS positions against
measured inter-drone geometry. We show that this relative-geometry channel has a
structural blind spot: a common, slowly varying translation (a \emph{rigid-covert}
shift, \rgd) preserves all pairwise distances and is therefore unobservable to
any relative-only detector (a gauge-freedom argument). We validate this blindness
on distance-verification and semidefinite-feasibility baselines, while explicitly
distinguishing it from onboard inertial/GNSS monitors that can raise a bare alarm
but cannot recover the swarm's true position.

To quantify when an external reference restores observability, we derive the
drift-dependent detection floor $2\gamma/(1-t_s/T)$ for a calibrated
anchor-residual detector and empirically identify an additional
detector-specific noise floor (measured slope $2.66$ vs.\ predicted $2.67$). We
then present a centralized anchor-rooted recovery pipeline that reconstructs
swarm geometry from inter-drone ranges, aligns it to a trusted-anchor
subset with Byzantine-robust fitting, and recovers the absolute positions of
non-anchored drones. A segmented estimator jointly estimates anchor drift, attack
rate, and onset when no clean-epoch label is available.

Across statistical simulations, ArduPilot software-in-the-loop experiments, and
Gazebo experiments with rendered vision anchors, the method recovers the positions
of non-anchored drones to a median error of $0.39$\,m ($20$ seeds) under
approximately $10.1$\,m of GNSS drift, and to $7.1$\,cm ($5$ seeds) in the
rendered-vision multi-SITL setting. We also characterize the explicit limits
imposed by non-collinear anchor geometry, anchor coverage, $\tau\!\to\!0$
drift--attack aliasing, and majority anchor compromise. All evaluations are
simulation-based and use no RF spoofing hardware or physical swarm.
\end{abstract}

\ccsdesc[500]{Security and privacy~Mobile and wireless security}
\ccsdesc[300]{Security and privacy~Embedded systems security}
\ccsdesc[300]{Computer systems organization~Robotic autonomy}

\keywords{GNSS spoofing; UAV swarm; cyber-physical security; sensor attack
detection; gauge freedom; cooperative localization}

\maketitle

\section{Introduction}\label{sec:intro}
\begin{figure*}[t]
\centering
\includegraphics[width=0.95\textwidth]{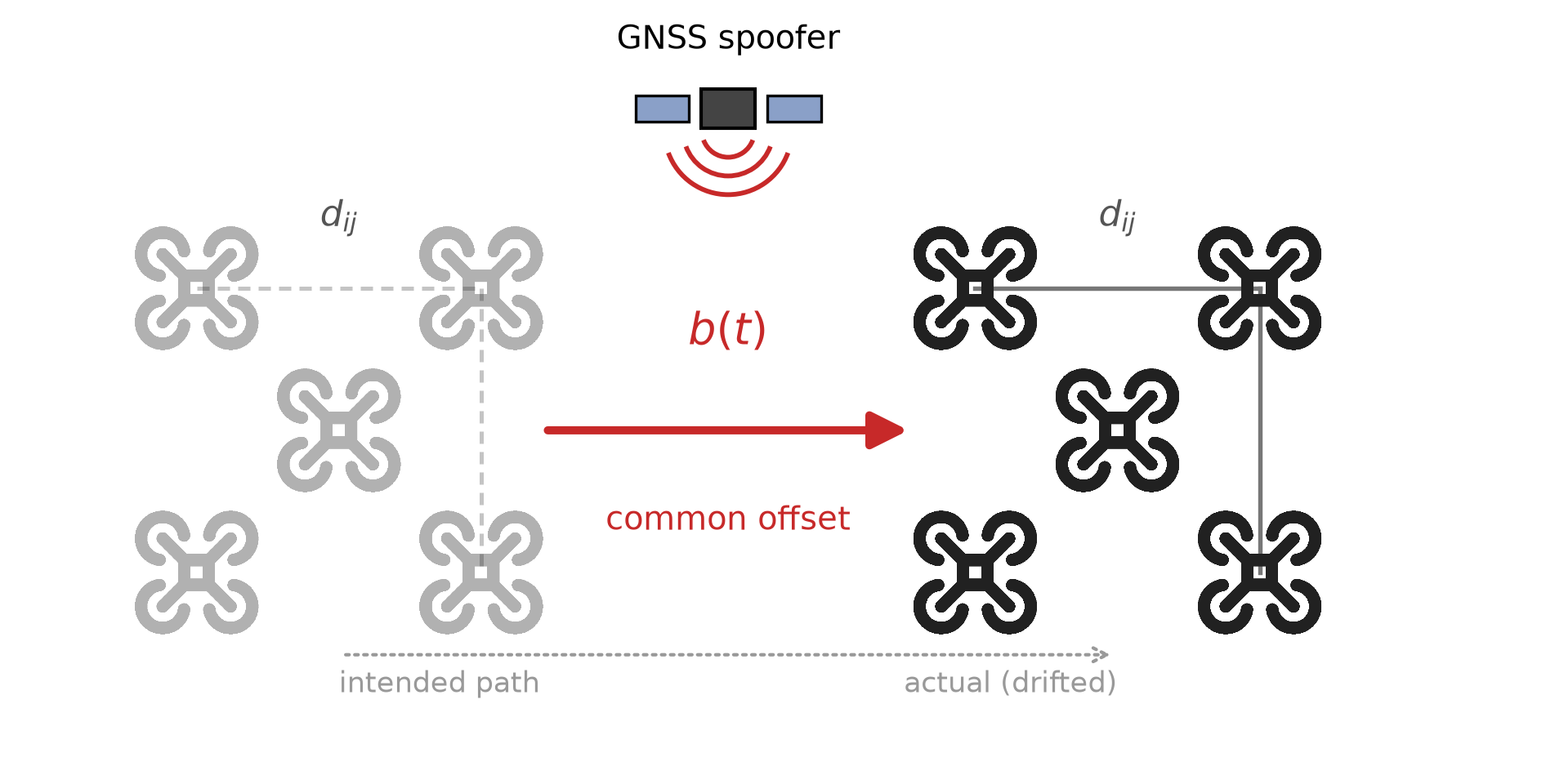}
\caption{\textit{RigidShift threat.} A coordinated GNSS spoofer adds a slow,
common offset $b(t)$ to every drone, so the whole formation drifts together as a
rigid body---shown solid at its actual drifted position, with the vacated intended
position as the faded ghost. Because every pairwise distance $d_{ij}$ is preserved,
a cooperative defense, which sees only relative quantities, finds no
inconsistency; being gradual, the shift also stays under each drone's built-in
glitch/failsafe gate. This is a structural blind spot.}
\label{fig:threat}
\end{figure*}
Autonomous UAV swarms are being deployed for delivery, inspection, agriculture,
and surveillance, and they depend on the Global Navigation Satellite System
(GNSS) for absolute positioning. GNSS is also vulnerable to spoofing---civilian signals carry no source authentication: a
software-defined radio can synthesize counterfeit signals that override the
genuine ones~\cite{humphreys2008,tippenhauer2011,psiaki2016}, and full
capture-and-control of an airborne vehicle by GPS spoofing has been demonstrated
in the field~\cite{kerns2014,noh2019}. Because a single spoofed drone is a
localized fault, the swarm-security literature answers with \emph{cooperative}
defenses: each drone compares its GNSS-reported position against a feasible
region built from its neighbors' reported positions and \emph{measured}
inter-drone ranges (UWB, RSSI, or optical flow), flags the inconsistent
minority, and reaches Byzantine consensus on the ``genuine''
positions~\cite{swarmraft,bi2023,bdgd,michieletto2023}.

\paragraph{The threat.}
We study an attack that the relative-geometry channel misses. A single or coordinated
attacker drives the GNSS of the \emph{entire} formation with an offset $b(t)$
that is (i) \textbf{common-mode}---nearly the same for all drones; (ii)
\textbf{gradual}---a slow ramp that stays within onboard innovation gates; and
(iii) \textbf{geometry-preserving}---all pairwise distances are unchanged. We
call this rigid-covert spoofing (\rgd; Fig.~\ref{fig:threat}). The underlying capability is not merely hypothetical: wide-area, approximately
formation-preserving spoofing has been demonstrated on SDR
hardware~\cite{gnsswasp}, and a formal analysis proves a coordinated spoofer can
exactly rigid-shift up to nine receivers, with bounded residual error
beyond~\cite{chen2025}; the exact common shift at larger swarm
scale is an idealized distributed-spoofer oracle we use to probe defense
scalability, not a validated wide-area attack (\S\ref{sec:eval}).

\paragraph{Why cooperative defenses miss it.}
A detector that keys on the inconsistency between GNSS-reported positions and
measured geometry is a function of \emph{relative} quantities only. Such a
function is invariant under a common translation of all reported positions: the
feasible region shifts by the same $b$, the ranges are unchanged, and the
residual stays $\approx 0$ (Section~\ref{sec:threat}, Prop.~1). This is the same
gauge freedom that makes anchor-free network localization observable only up to
a global rigid transform~\cite{aspnes2006}. We confirm the blindness both across
the literature (Table~\ref{tab:defenses}) and empirically: our reproduction of a
cooperative semidefinite-feasibility detector~\cite{bi2023} flags a common-mode
ramp at only $1.7\%$---below its $5\%$ false-alarm floor.

\paragraph{What we do \emph{not} claim.}
Recent multimodal swarm detectors~\cite{orbitguardnet,tristream} add a per-drone
inertial/GNSS Kalman-residual and signal-quality channel. That channel is
\emph{not} gauge-blind, and it can catch a covert ramp---but only above a
sensitivity floor, and it yields a bare alarm: the swarm learns that it is being
spoofed, but not where it actually is. Our own experiments confirm that a purpose-built \emph{offline}
monitor over the per-drone EKF raw-innovation sequence---an onboard, non-geometric detector in the spirit of~\cite{b1onboard}---detects
the covert ramp on $20/20$ Tier-2 attack runs at each tested rate ($2$--$20$\,cm/s)
(its held-out no-attack false alarm was $1/10$ runs, a wide finite-sample
estimate; App.~\ref{app:monitor})---even
though the autopilot's own \emph{built-in} glitch/failsafe gate stays silent
throughout (RQ3). Hence detection is possible \emph{in principle}; we do not claim
``only we can detect.'' The blind spot is specifically in the \emph{relative-geometry}
channel; and even a detection-only alarm is insufficient for restoring safe
navigation, since it provides no corrected absolute state---the swarm still does
not know where it actually is. Our
contribution is the layer beyond a blind alarm: \emph{recovery} of the true
positions of the whole swarm, the \emph{detection limit} of that ability, and the
\emph{failure envelope} of the anchor that provides it (and, as a by-product,
per-drone attribution under the distinct partial-spoof variant).

\paragraph{Contributions.}
\begin{itemize}
\item \textbf{Structural blind spot and quantitative boundary}
(\S\ref{sec:threat}--\S\ref{sec:limit}). We formalize a rigid common-mode GNSS
shift that is unobservable to relative-geometry channels---an instantiation of
network localization~\cite{aspnes2006} and CPS attack
identification~\cite{pasqualetti2013}---validate the blind spot on
representative cooperative baselines, and derive the drift-dependent detection
floor $2\gamma/(1-t_s/T)$ for a calibrated anchor-residual detector, with an
empirically measured detector-specific noise floor (slope $2.66$ vs.\ predicted
$2.67$).
\item \textbf{Anchor-rooted whole-swarm recovery}
(\S\ref{sec:anchorfail}--\S\ref{sec:defense}). We reconstruct the swarm geometry
from inter-drone ranges, align it to a trusted-anchor subset with
Byzantine-robust fitting, and recover the absolute positions of both anchored and
non-anchored drones. A segmented estimator jointly estimates anchor drift, attack
rate, and onset when no clean-epoch label is available.
\item \textbf{Multi-tier validation and explicit limits}
(\S\ref{sec:impl}--\S\ref{sec:eval}). Across statistical simulation, ArduPilot
software-in-the-loop, and Gazebo-rendered vision anchors, we evaluate detection,
recovery, Byzantine robustness, and non-anchor trust propagation, and identify
the limits imposed by anchor geometry, coverage, $\tau\!\to\!0$ drift--attack
aliasing, and majority anchor compromise.
\end{itemize}

\paragraph{Scope.}
Every result is in simulation across three tiers: a seed-fixed Python kinematic
swarm (Tier-1), ArduPilot software-in-the-loop with its EKF and controller in the
loop (Tier-2)~\cite{ardupilot}, and Gazebo with a rendered camera~\cite{koenig2004}.
We use no RF hardware and no physical swarm; we treat the vision renders and the
ArduPilot-in-the-loop flights as realism evidence for an otherwise abstracted
anchor channel, not as field validation (\S\ref{sec:limits}).

\section{Background and Threat Model}\label{sec:threat}

\subsection{System model}
A swarm of $N$ drones has (unknown) true horizontal positions
$x_i(t)\in\mathbb{R}^2$. Each drone reports a GNSS position
$z_i(t)=x_i(t)+b_i(t)+n_i(t)$, where $b_i$ is an attacker-injected offset and
$n_i\sim\mathcal{N}(0,\sigma_g^2 I)$ is receiver noise. Drones measure noisy
pairwise ranges $d_{ij}(t)=\lVert x_i-x_j\rVert+\varepsilon_{ij}$, with
$\varepsilon_{ij}\sim\mathcal{N}(0,\sigma_r^2)$, over an independent channel
(UWB) that the attacker cannot forge through GNSS. Each drone in a subset $\mathcal{A}$ of size
$\rho N$ carries an \emph{absolute anchor}: a GNSS-independent horizontal position
estimate $a_i(t)=x_i(t)+e_i(t)$, where $e_i$ is the anchor's own error---per-frame
position noise of scale $\sigma_a$ plus a slow registration drift of rate $\gamma$.
Table~\ref{tab:notation} summarizes the notation.

\begin{table}[t]
\caption{Notation.}
\label{tab:notation}
\small
\resizebox{\columnwidth}{!}{%
\begin{tabular}{@{}ll@{}}
\toprule
$N,\ \rho$ & swarm size; fraction carrying an absolute anchor \\
$x_i,\ z_i,\ a_i$ & true / GNSS-reported / anchor position of drone $i$ \\
$b(t),\ v$ & common-mode attack offset; ramp rate $\lVert\dot b\rVert$ \\
$d_{ij}$ & measured inter-drone range \\
$e_i(t),\ g_i(t)$ & anchor total error; estimated drift (\S\ref{sec:est}) \\
$\sigma_a,\ \gamma$ & anchor per-frame noise (m); worst-case drift rate (cm/s) \\
$t_s,\ t_d,\ T$ & attack onset; anchor-drift onset; horizon \\
$\tau\!=\!t_s{-}t_d$ & attack onset relative to drift (used in the $\tau\!\to\!0$ barrier) \\
$v^{\ast}$ & minimum detectable covert ramp rate (security index) \\
$f$ & number of Byzantine (actively compromised) anchors \\
$\varepsilon$ & radial formation-deformation rate (1/s); $\varepsilon{=}0$ is perfectly rigid \\
\bottomrule
\end{tabular}%
}
\end{table}

\subsection{Attacker model}
The attacker controls the GNSS signal reaching every drone and imposes a
coherent, slowly growing common offset $b(t)$ that preserves geometry. A single
spoofer can preserve the formation exactly for $N\le9$~\cite{chen2025}, whereas
preserving it at larger scale requires a wide-area multi-transmitter
spoofer~\cite{gnsswasp}; a single spoofer beyond $N\le9$ instead deforms the
formation, which the relative channel detects (\S\ref{sec:eval}). Direction is
the attacker's choice; magnitude ramps from onset $t_s$ at rate $v$. The attacker
may additionally compromise up to $f$ anchors, either by echoing the spoof into
the anchor channel or by physically corrupting the reference. The attacker does
\emph{not} control the honest anchors, the UWB ranging channel, or the
communication/aggregation infrastructure; these constitute the trusted computing base. The attacker knows the
defense and may adapt (e.g., ramp just below $v^{\ast}$, or align the attack with
a drifting anchor)---adaptive strategies we evaluate in \S\ref{sec:eval}.

\subsection{Observability structure}
Two propositions organize what is and is not possible, followed by two conditional
caveats. They are instantiations of standard theory, cited accordingly: the
propositions themselves are not new theorems. Our contributions are the
quantitative detector-specific limit (\S\ref{sec:limit}), the joint drift/attack
estimator (\S\ref{sec:est}), and the multi-tier system evaluation
(\S\ref{sec:anchorfail}--\ref{sec:eval}).

\begin{description}
\item[Prop.~1 (gauge freedom).] Any detector that is a function only of the
relative quantities $\{z_i-z_j\}$ and $\{d_{ij}\}$ is invariant under
$z_i\mapsto z_i+c$ for a common $c\in\mathbb{R}^2$. Hence a common-mode $b(t)$ is
unobservable from relative channels alone. \emph{(Anchor-free localization is
observable only up to a global rigid transform~\cite{aspnes2006}.)}
\item[Prop.~2 (anchor observability).] With at least one absolute anchor a
common translation becomes observable; recovering a general rigid motion that
includes rotation requires at least two.
\end{description}

\noindent\textbf{Two identifiability caveats.} Beyond these two propositions, the
trusted-anchor assumption rests on two \emph{conditional} facts, which we treat as
limitations and as motivation for the temporal estimator rather than as theorems.
\emph{Caveat~1 (conditional attack/fault separation):} an attack and an anchor
fault are distinguishable only when their measurement signatures are linearly
independent---a \emph{homogeneous} common-mode anchor fault is indistinguishable
from a common-mode GNSS shift, and separating them needs a heterogeneous absolute
modality (\S\ref{sec:eval}), a conditional instance of CPS attack
identification~\cite{pasqualetti2013}. \emph{Caveat~2 (memoryless-only aliasing):}
a drifting anchor can alias a covert attack so that a \emph{memoryless} anchor
residual stays flat, but the onset kink at $t_s$ is a change point that a
\emph{sequential} statistic (CUSUM~\cite{page1954}, GLR) still detects; true
inseparability holds only when the attack is simultaneous and isomorphic to the
drift ($\tau\!\to\!0$). This motivates the temporal estimator of \S\ref{sec:est}
for attacks with a resolvable change point ($\tau\!>\!0$) and defines its
fundamental $\tau\!\to\!0$ limit.

\section{Why Cooperative Defenses Fail}\label{sec:why}
Table~\ref{tab:defenses} surveys representative swarm GNSS-spoofing defenses.
The relative-geometry components of the examined cooperative defenses lack an
independent absolute horizontal reference; by Prop.~1 that channel is blind to a
common-mode geometry-preserving shift (the added inertial/signal channel of the
multimodal detectors is not gauge-blind---see below). \textbf{Distance
verification}~\cite{distverif2023} tests $|\,\lVert z_i-z_j\rVert -
d_{ij}|>\theta$, which is invariant under a common shift.
\textbf{SwarmRaft}~\cite{swarmraft} re-localizes flagged drones from
\emph{neighbor} positions and ranges alone, with no fixed ground
infrastructure or absolute reference. \textbf{BDGD}~\cite{bdgd} fuses RSSI distance checks with a
reputation vote that requires an honest majority. \textbf{Bi et
al.}~\cite{bi2023} solve a semidefinite feasibility problem tied to
self-reported positions, so a uniform offset gives $X=\hat X$, trivially
feasible. The two 2026 multimodal detectors,
\textbf{OrbitGuardNet}~\cite{orbitguardnet} and the \textbf{tri-stream}
network~\cite{tristream}, add a per-drone inertial/signal-quality channel that is
not gauge-blind. Like any covert-ramp detector such a channel will have its own
sensitivity floor (its observations, noise, and statistic differ from ours, so the
specific law of \S\ref{sec:limit} does not transfer to it), but---and this is our
point---it still performs no true-position recovery.
Notably, the method of \textbf{Michieletto et al.}~\cite{michieletto2023} uses no
independent absolute reference (RSS $+$ GNSS/IMU only, Table~\ref{tab:defenses})
and may therefore remain vulnerable to a
sufficiently slow common-mode shift (stealthy slow attacks provably evade \emph{any}
such model-based anomaly detector~\cite{khazraei2024}); and
\textbf{ASD/RSOM}~\cite{meng2023} explicitly leaves visual ranging to recover
the true location of spoofed drones to future work---the recovery problem we solve.

\begin{table}[t]
\caption{Representative swarm GNSS-spoofing defenses \emph{and} onboard detectors
vs.\ the rigid-covert threat. ``Abs.\ ref.'': independent absolute horizontal
reference; ``Blind'': blind to the rigid-covert threat; ``Recovery'': recovers true positions rather than only raising an alarm.
Detection is \emph{not} unique to us---a non-geometric monitor also detects---but
only anchor-rooted geometry \emph{recovers} the swarm.}
\label{tab:defenses}
\small
\resizebox{\columnwidth}{!}{%
\setlength{\tabcolsep}{4pt}\begin{tabular}{@{}p{2.75cm}p{1.55cm}ccc@{}}
\toprule
\textbf{Defense} & \textbf{Detection signal} & \textbf{Abs.\ ref.} & \textbf{Blind} & \textbf{Recovery} \\
\midrule
Distance verif.~\cite{distverif2023} & GNSS vs.\ UWB range & no & yes & no \\
Byz.\ dist.\ consensus~\cite{bdgd} & RSSI + reputation & no & yes & no \\
SDP feasibility~\cite{bi2023} & reported pos.\ + range & no & yes & no \\
Consensus reloc.~\cite{swarmraft} & peer pos.\ + range & no & yes & no \\
Multimodal~\cite{orbitguardnet,tristream} & +\,inertial/signal & no & geom.\ only & no \\
Formation loc.~\cite{michieletto2023} & RSS + GNSS/IMU & no & likely (rig.) & no \\
Offline EKF-innov.\ monitor (ours, cf.~\cite{b1onboard}; App.~\ref{app:monitor}) & onboard EKF innov. & no & no & no \\
\midrule
\textbf{This work} & \textbf{abs.\ anchor + ranges} & \textbf{yes} & \textbf{no} & \textbf{yes} \\
\bottomrule
\end{tabular}%
}
\end{table}

Appendix Fig.~\ref{fig:auc} makes the point quantitatively: both existing relative
detectors remain near chance across the sweep---distance verification peaks at only
$0.56$ AUC at $2$\,cm/s and both fall to $\approx0.48$--$0.49$ at $5$--$20$\,cm/s
(bootstrap $95\%$ CIs otherwise include $0.5$)---while our over-determined-geometry and
absolute-anchor detectors reach $1.0$ above the detection limit.

\section{The Detection-Limit Law}\label{sec:limit}
\begin{figure}[t]
\centering
\includegraphics[width=\columnwidth]{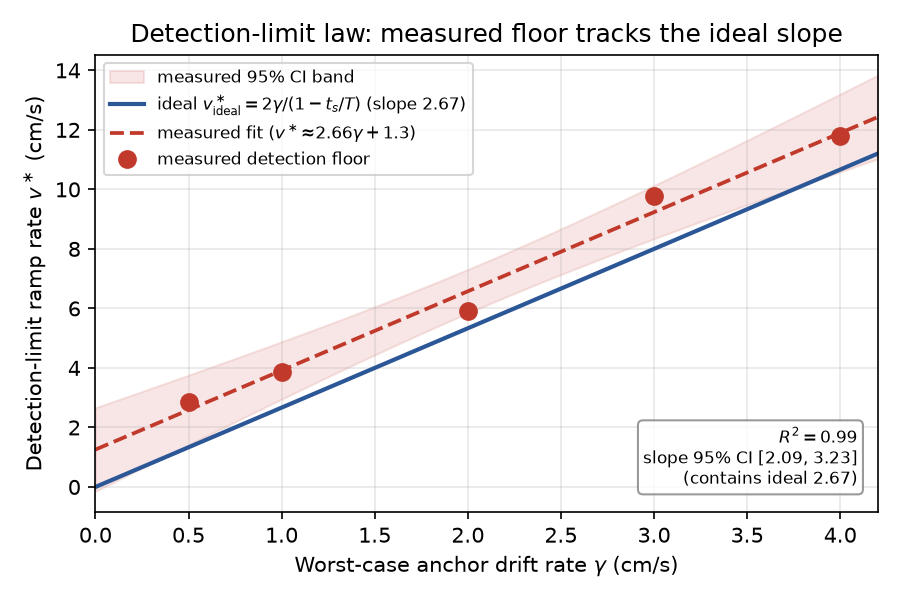}
\caption{Detection-limit law validation: the measured floor tracks the derived
slope of Eq.~\eqref{eq:vstar} (measured $2.66$ vs.\ predicted $2.67$; fit
$R^2{=}0.99$, slope $95\%$ CI $[2.09,3.23]$ over $n{=}5$ drift levels, containing
the predicted slope), with a
positive intercept ($\approx\!1.3$\,cm/s) consistent with a per-frame-noise floor
$v_{\text{noise}}$ (at $\sigma_a{=}0.5$\,m) that the idealized law omits. $\gamma$
is the worst-case anchor-drift rate.}
\label{fig:vstar}
\end{figure}

Within our recovery schema, the independent anchor is the absolute reference
outside the relative gauge, so an anchor-referenced detector bounds what a
relative-blind defense can add. We derive the floor for a worst-case
covert attacker against a single anchor modality whose registration slowly drifts.
The attacker ramps the GNSS offset at rate $v$ from onset $t_s$; independently, the
honest anchor's registration drifts at rate $\gamma$ (the modality's worst-case
drift capability, in cm/s), and in the worst case this drift is \emph{aligned}
with the attack direction. The attacker anticipates or observes the drift but does
\emph{not} control the honest anchor---an attacker-controlled anchor is the
separate Byzantine case of \S\ref{sec:defense}. The median anchor residual is then
$r(t)=\lvert v\,(t-t_s)_+ - \gamma\,t\rvert$, and a detector calibrated at the
false-alarm quantile of the drift-only run (end-of-horizon peak $\sim\gamma T$)
fires only when the attack's residual peak clears that band,
$(v-\gamma)T - v\,t_s > \gamma T$, giving the ideal detection floor
\begin{equation}
v^{\ast}_{\text{ideal}} \;=\; \frac{2\gamma}{\,1-t_s/T\,}.
\label{eq:vstar}
\end{equation}
Both sides are rates. A realistic detector also faces the anchor's per-frame
position noise $\sigma_a$, which imposes a noise floor
$v_{\text{noise}}(\sigma_a,T,\alpha)$ at operating point $\alpha$ that persists
even for a non-drifting modality, so the operational security index is
\begin{equation}
v^{\ast} \;=\; \frac{2\gamma}{\,1-t_s/T\,} \;+\; v_{\text{noise}}(\sigma_a,T,\alpha).
\label{eq:vstar2}
\end{equation}
Sweeping $\gamma$ at fixed $\sigma_a{=}0.5$\,m, the measured floor is
$v^{\ast}\!\approx\!2.66\,\gamma + 1.3$ (cm/s): the slope matches the derived
$2/(1-t_s/T)=2.667$ ($t_s/T{=}0.25$; fit $R^2{=}0.99$, slope $95\%$ CI
$[2.09,3.23]$ contains it), and the $1.3$\,cm/s intercept is consistent
with a per-frame-noise floor $v_{\text{noise}}$ at these settings
(Figure~\ref{fig:vstar}). Eq.~\eqref{eq:vstar2} is therefore an \emph{operational}
(empirical) decomposition: we verify the $\gamma$-linear term and an additive floor
at $\sigma_a{=}0.5$\,m, $t_s/T{=}0.25$; the dependence of $v_{\text{noise}}$ on
$\sigma_a$, $T$, and the false-alarm point $\alpha$---and the linearity of the split
under correlated or heavy-tailed noise---are not separately derived. A drift-free
modality ($\gamma\!\to\!0$, e.g.\ a fixed UWB beacon) drives the floor down to
$v_{\text{noise}}$, not to zero, while a drift-prone modality (vision) raises it
with $\gamma$. We present it as a calibrated-detector security index for this
threat model, not a bound over every conceivable detector.

\paragraph{Adaptive attacker / security index.}
An attacker who ramps at a \emph{constant rate} just below $v^{\ast}$ accumulates a
covert displacement budget $v^{\ast}\!\cdot\!(T-t_s)$; thus $v^{\ast}$ is a security
index \emph{for the constant-rate ramp family}. Nonlinear profiles evade the same
detector longer---a back-loaded polynomial delays time-to-detect $5.7\times$
(\S\ref{sec:eval})---and need a separate displacement- or energy-constrained
analysis, so we do not claim $v^{\ast}(T{-}t_s)$ as a budget for arbitrary
profiles. With $K$ independent modalities of drift $\gamma_j$ and noise floor
$v_{\text{noise},j}$, a type-aware detector keying off the most favorable modality
gives $v^{\ast}(K)=\min_j\!\big[2\gamma_j/(1-t_s/T)+v_{\text{noise},j}\big]$.
Appendix Fig.~\ref{fig:secidx} bears the trend out---adding clean references drives
$v^{\ast}$ down ($10\!\to\!6$\,cm/s realistic, to $2$ for an oracle)---though the
realistic detector needs a clean \emph{majority} to trust the clean modality, and
fusing several clean references lowers the floor beyond the single-modality
$\min$ (oracle $3\!\to\!2$); the exact multi-modality relationship is empirical.

\paragraph{Relation to prior limits.}
Khazraei et al.~\cite{khazraei2024} prove that a stealthy GPS attack can cause
unbounded deviation on a single vehicle, and Murguia \& Ruths~\cite{murguia2017}
bound the state degradation an undetected attacker can induce on an LTI CPS.
Both are existence/impact statements, not a drift-proportional ramp-rate
threshold against an external absolute anchor in a swarm. The closest analog,
Baweja~\cite{baweja2026}, proves a structurally similar slow-ramp impossibility
in the single-receiver \emph{timing/clock} domain; Eq.~\eqref{eq:vstar} is
position/swarm/anchor-specific and we distinguish it explicitly.

\section{Anchor Failure Modes}\label{sec:anchorfail}
The absolute anchor is necessary but not automatic: it introduces its own failure
envelope. Table~\ref{tab:anchorfail} summarizes four modes we characterize
empirically in the rigid-covert swarm setting; the detailed sweeps are in
Appendix~\ref{app:supp}.

\begin{table}[t]
\caption{Anchor failure modes (rigid-covert swarm; detailed sweeps in
Appendix~\ref{app:supp}).}
\label{tab:anchorfail}
\small
\begin{tabular}{@{}p{1.7cm}p{3.15cm}p{2.4cm}@{}}
\toprule
\textbf{Mode} & \textbf{Observed effect} & \textbf{Design implication} \\
\midrule
Aligned drift & Memoryless residual masked ($1.0\!\to\!0.0$ at aligned drift
$\approx v/2$); sequential CUSUM/GLR still catch the onset (AUC $1.0$) &
Temporal drift/attack separation (\S\ref{sec:est}) \\
Coverage loss & Vision tracking $\approx\!6$\,cm within a $\approx\!4$\,m
envelope, then diverges & Coverage is a deployment constraint \\
Correlated failure & Common-mode anchor errors collapse detection
($1.0\!\to\!0.17$ at $10\%$ co-failure) & Heterogeneous independent modalities \\
Dilution & Fixed anchor count: far-node error grows with $N$
($14.5\!\to\!37.2$\,m, $N{=}16\!\to\!128$, $3$ anchors) & Anchor ratio and
placement govern recovery \\
\bottomrule
\end{tabular}
\end{table}

Absolute anchors break the relative gauge but introduce their own failure
envelope. Aligned anchor drift can mask a \emph{memoryless} residual (detection
$1.0\!\to\!0.0$ at an aligned drift $\approx v/2$; App.~Fig.~\ref{fig:aliasing}),
but a $320$-configuration enumeration confirms that only memoryless/peak detectors
are masked---sequential CUSUM/GLR still catch the onset kink---so aliasing
motivates the temporal estimator of \S\ref{sec:est} for \emph{recovering} attacks
with a resolvable change point ($\tau\!>\!0$) rather than for detection; the
simultaneous isomorphic case $\tau\!\to\!0$ remains unidentifiable
(\S\ref{sec:barrier}). Landmark coverage
limits vision-based anchors (Figure~\ref{fig:swarm8}b); correlated common-mode
anchor errors defeat redundancy (App.~Fig.~\ref{fig:anchorfail}a); and a fixed
anchor count produces geometric dilution as the swarm grows
(App.~Fig.~\ref{fig:anchorfail}b), so both the anchor \emph{ratio} and the
non-collinear \emph{placement} (\S\ref{sec:arch}) govern recovery. We treat these
as deployment and identifiability conditions rather than separate contributions.

\section{Defenses}\label{sec:defense}
\begin{figure*}[t]
\centering
\includegraphics[width=0.95\textwidth]{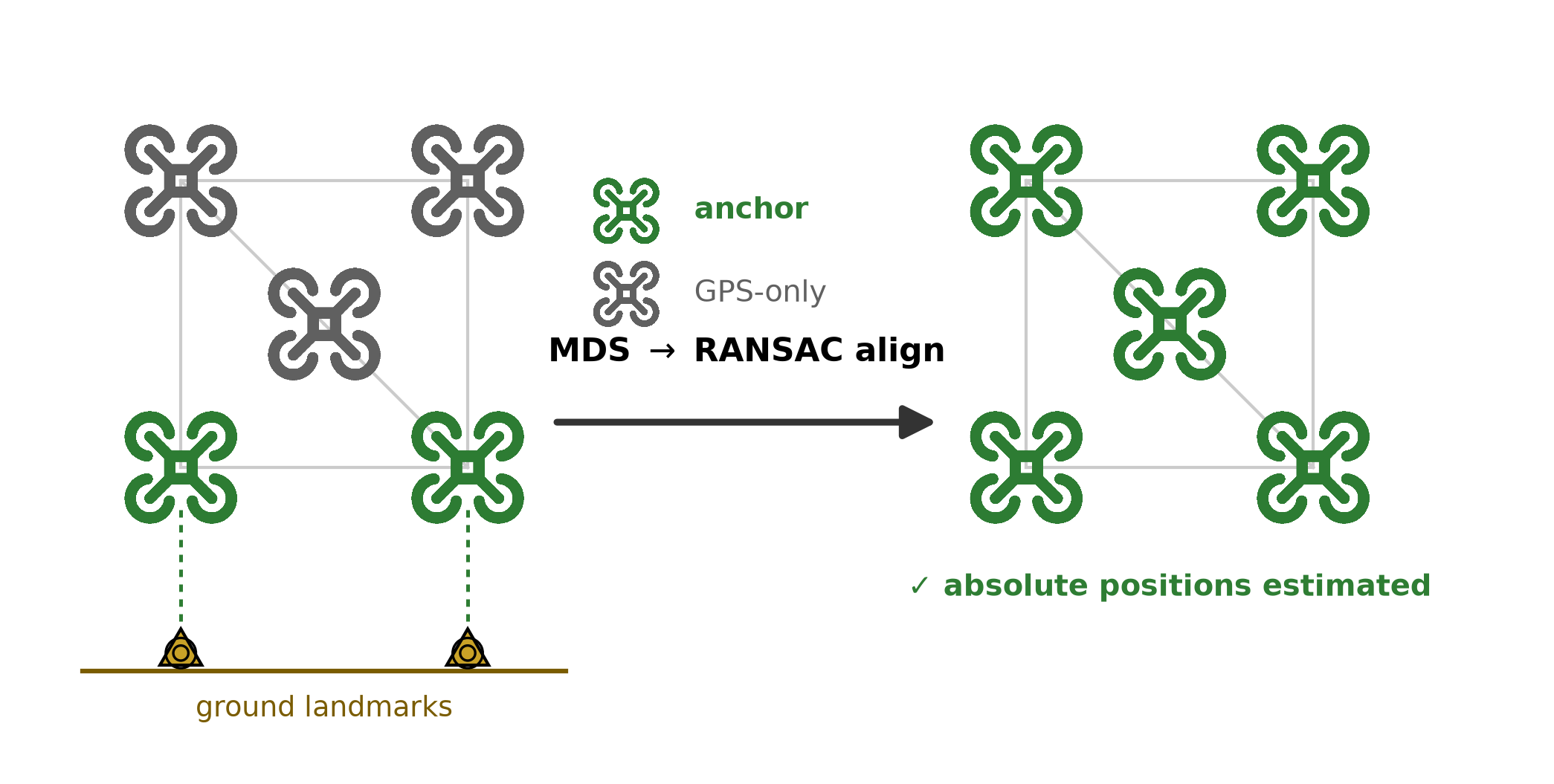}
\caption{\textit{Anchor-rooted recovery.} A few drones carry an anchor (green):
they see a fixed ground landmark, giving them a GNSS-independent estimate of their
true position that a GNSS spoof cannot alter. The inter-drone ranges $d_{ij}$ fix the swarm's shape but not its
absolute placement (the gauge freedom of Fig.~\ref{fig:threat}); aligning that
shape to the trusted anchors (classical MDS, then RANSAC) estimates the absolute
position of every drone, including the GPS-only ones.}
\label{fig:defense}
\end{figure*}
The defense has three components: the anchor-rooted recovery architecture with optional
attribution (\S\ref{sec:arch}; Fig.~\ref{fig:defense}) and the two temporal estimators (bootstrap and joint) that keep it working when the
anchor itself drifts (\S\ref{sec:est}). All three consume a common schema
(position, range, anchor, timestamp), so identical code runs on synthetic and
simulator data (\S\ref{sec:impl}).

\subsection{Anchor-rooted whole-swarm recovery}\label{sec:arch}
A subset $\mathcal{A}$ ($\rho N$ drones) carries an absolute anchor. The recovery
reconstructs the swarm shape from the
over-determined ranges by classical multidimensional scaling (MDS), robustly
fits the alignment onto the trusted anchors under the evaluated
Byzantine-minority conditions (RANSAC~\cite{fischler1981}), and propagates the single anchored frame to every
non-anchor drone. Recovery is thus a single \emph{centralized} global estimate;
robustness to Byzantine anchors comes from the RANSAC alignment, not from a
distributed consensus (a fully distributed realization would replace this with an
iterated W-MSR protocol~\cite{leblanc2013}, which we do not implement or evaluate).
Attribution flags drone $i$ as spoofed when its GNSS report disagrees with its
recovered position, $\lVert z_i-\hat x_i\rVert>\theta$.

The recovery is three standard steps on the per-frame ranges $D{=}\{d_{ij}\}$ and
trusted anchors $\{a_i\}_{i\in\mathcal{A}}$:
\begin{equation}\label{eq:recover}
\begin{aligned}
Y &= \textsc{ClassicalMDS}(D),\\
(\hat R,\hat t) &= \textsc{RANSACAlign}(Y_{\mathcal{A}},\{a_i\}_{\mathcal{A}}),\\
\hat X &= \hat R\,Y + \hat t.
\end{aligned}
\end{equation}
The first step reconstructs the swarm shape up to a rigid transform; the second
aligns it to the trusted anchors while rejecting a Byzantine minority of $\le f$
anchors; the final transform recovers all $N$ absolute positions $\hat X$,
including drones with no anchor of their own.

\noindent\textbf{Assumptions and minimum conditions.} Classical MDS requires a
\emph{complete} distance matrix; our large-$N$ sweeps supply it with all-pairs UWB,
and a geodesic range-graph-completion step (which needs a \emph{connected} graph)
extends recovery to a sparse graph---robust down to
$\approx\!35\%$ density before connectivity fails (\S\ref{sec:eval},
App.~Fig.~\ref{fig:sparse})---and heavy non-line-of-sight tails are handled robustly
(evaluated on a real UWB dataset, \S\ref{sec:eval}). We model the horizontal plane only
(the covert offset is horizontal; altitude is barometric and outside the threat
model). The anchor counts separate cleanly: \emph{detecting} a common translation
needs one trusted anchor (Prop.~2); fixing the \emph{orientation} (a rotation
$R\in SO(2)$) needs two; and \emph{removing} the MDS reflection ambiguity for full
absolute recovery needs three trusted anchors whose true coordinates are
\emph{non-collinear} (two points always lie on a line and cannot resolve the
reflection about it; we therefore use non-collinear anchors in the multi-SITL
capstone, \S\ref{sec:eval}). Rejecting $f$ Byzantine anchors by RANSAC therefore needs both a
trusted majority and three non-collinear honest inliers, i.e.\ $m\ge\max(2f{+}1,\,
f{+}3)$ present anchors, where $m$ denotes the number of available anchors. Our recovery is \emph{centralized} (a single MDS $+$
RANSAC estimate), which we implement and evaluate; a fully \emph{distributed}
realization would add an iterated W-MSR consensus over a
$(2f{+}1)$-robust communication graph against an $f$-local
adversary~\cite{leblanc2013}, which we leave as a deployment extension rather than
claim as evaluated.

\subsection{Drift-learning estimators}\label{sec:est}
When the anchor drifts, a covert ramp aligned to the drift can pass the anchor
residual (Caveat~2). Both estimators exploit \emph{time}: a benign drift is a
smooth process, whereas an attack is a change-point ramp.

\noindent\textbf{Bootstrap drift-correction.} Given a clean (no-attack) epoch,
fit each anchor's drift model $\hat g_i(t)$, extrapolate it forward, and subtract
it from the anchor residual; what remains is the attack. An attack that aliases
against the raw anchor is exposed against the drift-corrected anchor.

\noindent\textbf{Joint estimator.} The clean-epoch label is itself an
assumption. The joint estimator (Algorithm~\ref{alg:joint}) drops it: by
segmented (change-point) regression it fits, over the whole record and
\emph{per axis}, a model in which the signal is a continuous anchor drift $g$ plus
an attack ramp of rate $v$ starting at an unknown onset $t_s$, and returns
$(\hat g,\hat v,\hat t_s)$. This resolves the \emph{instantaneous} drift--attack
aliasing by using the \emph{temporal} signature of the change point---it does
\emph{not} lift the $50\%$ Byzantine barrier (\S\ref{sec:defense})---and it is
well-posed only when the pre-onset window is long enough to fit the drift
(\S\ref{sec:eval}); with too little clean history the
drift and attack are collinear and the estimate degrades.

\begin{algorithm}[t]
\caption{Joint drift/attack estimator over the aggregated anchor residual}
\label{alg:joint}
\begin{algorithmic}[1]
\Require per-axis residual $r_c(t)$ (coordinate-wise median over usable anchors),
      $c\in\{x,y\}$; drift basis
      $\Phi(t)=[1,\,t]$; onset grid $t_s\in\{5,6,\dots,59\}$\,s
\For{each candidate onset $t_s$ in the grid}
  \State per axis $c$: fit $r_c(t)\approx \Phi(t)\beta_c + v_c\,(t-t_s)_+$ (LS)
  \State $\text{RSS}(t_s)\gets \sum_c$ residual sum of squares
\EndFor
\State $\hat t_s \gets \arg\min_{t_s} \text{RSS}(t_s)$;\ \
      $(\hat\beta_c,\hat v_c)\gets$ fit at $\hat t_s$
\State $\hat v \gets \lVert(\hat v_x,\hat v_y)\rVert$ \Comment{magnitude; direction $=\angle(\hat v_x,\hat v_y)$}
\State \Return $\hat g=\Phi\hat\beta,\ \hat v,\ \hat t_s$
\end{algorithmic}
\end{algorithm}

\noindent\textbf{Details.} The residual is $r_c(t)=\operatorname{median}_i
(z_{i,c}-a_{i,c})$ over usable anchors, per axis; the two axes are fit
independently, so the attack \emph{direction} is recovered as
$\angle(\hat v_x,\hat v_y)$, not assumed. The drift basis is affine ($[1,t]$),
matching a slowly-registering anchor; the onset grid is $1$\,s. The fit is
$O(|\text{grid}|\cdot T)$ for the single aggregated residual stream. We report accuracy as the mean
absolute error over $40$ seeds per condition; the ranges in
Table~\ref{tab:defres} are over drift rate $\gamma\!\in\![0,3]$\,cm/s and onset
$t_s\!\in\![20,40]$\,s (accuracy holds for pre-onset windows $\gtrsim\!10$\,s and
degrades below, as drift and attack become collinear).

\subsection{Fundamental barriers}\label{sec:barrier}
Two \emph{independent} barriers remain; either alone defeats this defense family,
and neither estimator beats them. \emph{First (Byzantine majority):} active
compromise of an anchor majority ($f\ge m/2$) makes the trusted transform
ambiguous---the trusted set no longer out-votes the compromised set. \emph{Second
(temporal identifiability):} even with fully honest anchors, an attack that is
simultaneous and isomorphic to the anchor drift ($\tau\!\to\!0$) is unidentifiable
under our measurement model, because it leaves no change point for the temporal
estimator to key on. Robustness degrades gracefully toward each wall
(\S\ref{sec:eval}); we state them as limits, not gaps.

\section{Implementation and Setup}\label{sec:impl}
We use three tiers with a shared detector schema so identical detector code runs
on all of them. \textbf{Tier-1} is a seed-fixed \textsc{numpy}/\textsc{scipy}
kinematic swarm (seed-fixed, dependency-pinned reproduction), used for statistical sweeps.
\textbf{Tier-2} is ArduPilot software-in-the-loop (SITL)~\cite{ardupilot} with
its EKF3 and flight controller in the loop; the covert ramp is injected through the
simulator GPS glitch parameter \texttt{SIM\_\allowbreak GPS1\_\allowbreak GLTCH}, an L1-sensor injection
point upstream of the EKF. \textbf{Gazebo}~\cite{koenig2004} renders a downward
camera over five known-position coloured landmarks; the vision anchor recovers
absolute position by HSV colour segmentation and \texttt{solvePnP}. Experiment
configurations are declarative and versioned; the artifact (detector code,
configs, and the swarm harness) is publicly available (Appendix~\ref{app:repro}). Threat-model contract: detectors read only GNSS positions, ranges,
and anchor positions; ranges and anchors come from an independent channel the
attacker cannot forge through GNSS. Table~\ref{tab:realism} delineates which signals are genuine simulator outputs
versus modeled channels per tier, and Table~\ref{tab:params} lists the default
parameters; sweeps vary one axis at a time around them.

\paragraph{Statistics.} Unless noted, the bootstrap resampling unit is one
\emph{independent run}: for Tier-1, a fresh seed drives the whole kinematic sim
(trajectory, GNSS, range, and anchor noise); for Tier-2, one SITL flight plus an
independently drawn range/anchor-noise realization. Some Tier-2 CIs are narrow
because recovery error is a geometry-dominated statistic averaged over the swarm
and horizon, so its run-to-run spread is genuinely small---we confirmed this is
not a fixed-noise artifact by re-drawing the per-run noise seed, which left the
means and CIs essentially unchanged. Where a mechanism is discrete (e.g.\ RANSAC
inlier-selection failure at the $50\%$ Byzantine boundary) the CI is correspondingly wider.

\begin{table}[t]
\caption{Default simulation parameters (Tier-1 unless noted). Sweeps vary one
axis around these defaults.}
\label{tab:params}
\small
\begin{tabular}{@{}llll@{}}
\toprule
\textbf{Parameter} & \textbf{Value} & \textbf{Parameter} & \textbf{Value} \\
\midrule
Swarm size $N$        & $8$ ($4$--$128$) & GNSS noise $\sigma_g$   & $0.50$\,m \\
Formation / spacing   & grid / $5$\,m    & Range noise $\sigma_r$  & $0.10$\,m \\
Horizon $T$ / step    & $80$\,s / $0.1$\,s & Anchor noise $\sigma_a$ & $0.50$\,m \\
Cruise speed          & $3$\,m/s         & Anchor ratio $\rho$     & $0.5$ ($0$--$0.5$) \\
Onset $t_s/T$         & $0.25$           & Covert ramp $v$         & $2/5/20$\,cm/s \\
Seeds / condition     & $200$ (AUC)      & Attribution thr.\ $\theta$ & $1$\,m \\
\bottomrule
\end{tabular}
\end{table}

\begin{table}[t]
\caption{What is a genuine simulator output vs.\ a modeled channel, per tier. No
tier transmits or receives an RF signal.}
\label{tab:realism}
\small
{\hyphenpenalty=10000\exhyphenpenalty=10000\tolerance=9999
\resizebox{\columnwidth}{!}{%
\begin{tabular}{@{}p{1.9cm}p{1.35cm}p{2.1cm}p{2.1cm}@{}}
\toprule
\textbf{Component} & \textbf{Tier-1} & \textbf{Tier-2} & \textbf{Vision capstone} \\
\midrule
Vehicle / controller & kinematic & ArduPilot SITL & ArduPilot $+$ Gazebo \\
GNSS attack & software inject & sim.\ GPS-glitch & sim.\ GPS-glitch \\
Inter-drone ranges & synthetic & truth-derived $+$ noise & truth-derived $+$ noise \\
Absolute anchor & synthetic & synthetic / noisy & rendered image pipeline \\
RF signal & none & none & none \\
\bottomrule
\end{tabular}%
}}
\end{table}

\paragraph{Detector definitions.} Every detector reduces each frame to a scalar
score, \emph{gates} it (fires only after $\ge3$ consecutive frames---$0.3$\,s at
$dt{=}0.1$---above threshold), and calibrates the threshold at the $\alpha{=}5\%$
false-alarm quantile of $120$ no-attack runs; the ROC/AUC score is the gated
run-peak. The two \emph{relative} baselines score geometry inconsistency:
\emph{distance verification}~\cite{distverif2023} on
$\max_{ij}\lvert\lVert z_i-z_j\rVert-d_{ij}\rvert$, and the \emph{cooperative SDP}
on the reported-position/range feasibility residual. Our \emph{absolute anchor}
detector scores the median over anchored drones of $\lVert z_i-a_i\rVert$; our
\emph{over-determined geometry} detector scores the median over \emph{all} drones
of $\lVert z_i-\hat x_i\rVert$, where $\hat X=\textsc{MDS}(\{d_{ij}\})$ rigidly
aligned to the anchors (Eq.~\eqref{eq:recover}), falling back to the anchor-only
residual when fewer than three anchors are present. The two differ only in whether
trust is propagated to non-anchored drones through the range geometry.

\paragraph{Robust RANSAC alignment.} The anchor alignment
(\textsc{RANSAC\-Align}, Eq.~\eqref{eq:recover}) fits a rigid transform from a
minimal $3$-anchor sample and, rather than sampling randomly, \emph{exhaustively}
enumerates all $\binom{m}{3}$ anchor triples (so the estimate is deterministic---
there is no iteration count or success-probability trade-off). Each candidate is
scored by its inlier set with an inlier threshold of $1.5$\,m; ties on inlier
count are broken by the smallest total inlier residual (an MSAC-style rule that
avoids a triple containing a Byzantine anchor); the winning transform is then re-fit
(least-squares Procrustes) on its full inlier set. With fewer than four anchors,
or if no triple yields $\ge3$ inliers, it falls back to plain Procrustes. The
exhaustive $O(m^3)$ enumeration is the source of the large-$N$ recovery cost in
Appendix Table~\ref{tab:runtime}. All parameters and the code are in the artifact.

\section{Evaluation}\label{sec:eval}
We organize the evaluation around three questions. \textbf{RQ1} asks whether
representative relative-geometry defenses are blind to the rigid common shift and
whether the measured anchor detection floor follows the derived law. \textbf{RQ2}
asks whether anchor-rooted recovery reconstructs both anchored and non-anchored
drones under drift, sparse ranging, and Byzantine anchors. \textbf{RQ3} asks
whether the complete defense loop holds in multi-vehicle ArduPilot SITL with
Gazebo-rendered vision anchors, and where its explicit limits appear. Detailed
curves and secondary experiments are in Appendix~\ref{app:supp}.

\subsection{RQ1: Structural blindness and detection boundary}
\paragraph{Blindness and detection.}
Table~\ref{tab:auc} reports detection AUC under the common-mode attack over $200$
seeds/condition. Both existing relative detectors stay at chance (bootstrap
$95\%$ CIs include $0.5$); our detectors saturate at $1.0$ once the ramp exceeds
$v^{\ast}$ (App.~Fig.~\ref{fig:auc}). We reproduce the cooperative SDP-feasibility
class~\cite{bi2023} both as a dependency-free MDS$+$Procrustes proxy and as a true
\textsc{cvxpy} semidefinite program; both are relative-only, hence gauge-blind. At
the covert rates ($5$--$20$\,cm/s) the exact \textsc{cvxpy} SDP flags only $1/60$
attack runs ($1.7\%$), at the $5\%$-false-alarm operating point. Because the exact
SDP is a per-frame binary feasibility test rather than a graded score, Table~\ref{tab:auc}
reports the cheaper proxy for the full multi-rate AUC sweep; the two agree
qualitatively (both at chance). The EKF-innovation monitor is excluded from this
Tier-1 table because it keys on per-drone EKF innovations available only in the
Tier-2 SITL runs; its quantitative detection results appear in
App.~\ref{app:monitor}.

\begin{table}[t]
\caption{Detection AUC under the common-mode covert attack (Tier-1, $200$
seeds/condition). Both existing detectors stay near chance (bootstrap $95\%$ CIs
include $0.5$, except distance verification at $2$\,cm/s, $0.56$;
App.~Fig.~\ref{fig:auc}); the exact \textsc{cvxpy} SDP independently confirms this
blindness, flagging the covert ramp on only $1.7\%$ of runs.}
\label{tab:auc}
\small
\begin{tabular}{@{}lccc@{}}
\toprule
\textbf{Detector} & \textbf{2 cm/s} & \textbf{5 cm/s} & \textbf{20 cm/s} \\
\midrule
Distance verification~\cite{distverif2023} (existing) & 0.56 & 0.49 & 0.48 \\
Cooperative SDP~\cite{bi2023} (proxy) & 0.54 & 0.49 & 0.48 \\
Over-determined geometry (ours)  & 0.84 & 1.00 & 1.00 \\
Absolute anchor (ours)           & 0.73 & 1.00 & 1.00 \\
\bottomrule
\end{tabular}
\end{table}

\paragraph{Detection floor and scaling.}
The measured detection floor tracks Eq.~\eqref{eq:vstar} at slope $2.660$
vs.\ predicted $2.667$ (Figure~\ref{fig:vstar}). Dense $(\rho,N)$ sweeps confirm
$v^{\ast}$ falls as $\rho$ grows: a $2$-anchor/$\rho{=}0.25$ configuration reaches
detection AUC $0.999$, and the anchor channel does not dilute out even at
$N{=}128$ (AUC $1.0$), given a connected range graph. Increasing $\rho$ does not
change the derived drift term of Eq.~\eqref{eq:vstar}; it lowers the
\emph{empirical} noise floor by aggregating multiple anchors through the median.
(These are \emph{detection} results; full whole-swarm \emph{recovery} additionally
requires three non-collinear trusted anchors, \S\ref{sec:arch}.) We are precise
about \emph{which attacker}
makes an exact common-mode shift feasible at scale. Exact covert blindness is
structural only to $\varepsilon{=}0$ (Prop.~1); any deformation
$\varepsilon\ge0.002$ self-reveals through the relative channel, a practical
threshold that \emph{tightens} with $N$ (App.~Fig.~\ref{fig:frontier}).

\noindent\emph{Is a near-exact shift a realistic strawman?} We ground the
attack's imperfection in GNSS-WASP's own hardware measurement---a $0.97$\,m
average relative-distance error at $\approx\!1000$\,m from the spoofer
reference~\cite{gnsswasp}---and inject that same distance-proportional residual at
swarm scale ($5$\,m spacing). At the real measured ratio, and up to
$\approx\!100\times$ it, the relative distance-check baseline stays at its
false-alarm floor (detection $0.07$ at the measured level), because at metre-scale
spacing the induced differential displacement is millimetric; our anchor detector
flags it throughout ($1.0$), and the relative channel begins to see the residual
only near $\approx\!500\times$ the measured ratio (App.\ Table~\ref{tab:secondary}).
So the covert-blind regime is not an artifact of exact $\varepsilon{=}0$: under this
distance-proportional extrapolation of the GNSS-WASP measurement, the resulting
swarm-scale residual is far below what the relative channel can exploit.

A single-antenna spoofer can preserve a
formation exactly only for $N\le9$~\cite{chen2025} (the multi-receiver
countermeasure of \cite{tippenhauer2016} exploits exactly this single-transmitter collapse~\cite{chen2024single}), so
preserving a formation across a large swarm requires a \emph{wide-area,
multi-transmitter} spoofer. The $N{=}128$ case evaluates defense scalability under
an \emph{ideal distributed-spoofer oracle}---a perfectly common offset injected in
software. GNSS-WASP~\cite{gnsswasp} establishes wide-area formation-preserving
feasibility, but we do not reproduce its transmitter geometry, inter-transmitter
synchronization, or receiver-level residuals at $N{=}128$, and do not claim those
residuals stay below our detection threshold there.

\subsection{RQ2: Whole-swarm recovery and robustness}
\begin{figure}[t]
\centering
\includegraphics[width=\columnwidth]{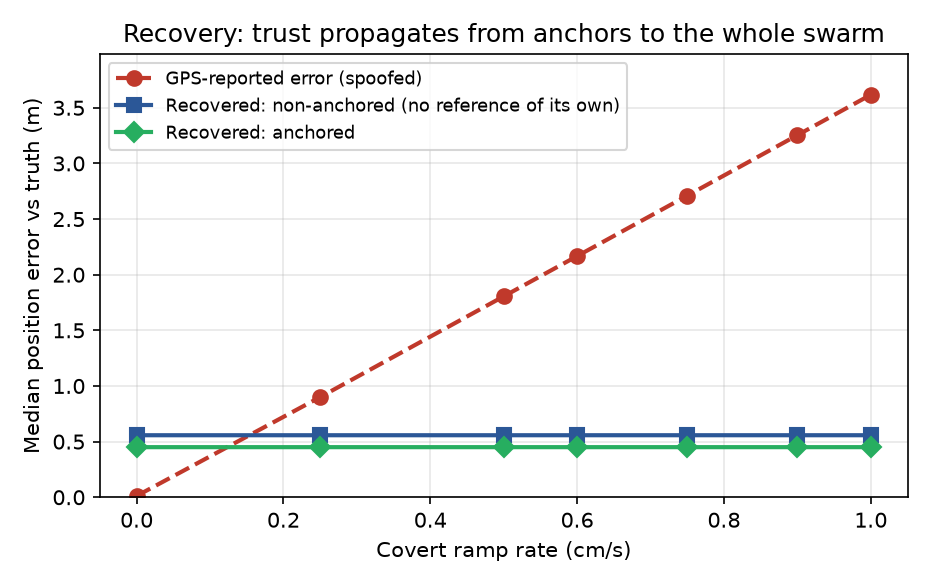}
\caption{Recovery and cooperation gain (low-rate $0$--$1.0$\,cm/s, $400$\,s Tier-2 SITL harvest, ArduPilot EKF-in-the-loop; $10$
seeds/rate). Reported-vs-true error ramps with the covert attack, while
anchor-rooted recovery holds both anchored ($0.45$\,m) and non-anchored
($0.56$\,m) drones near the anchor-noise-limited floor ($\approx\!0.5$\,m)---trust propagates from the anchor
subset to drones with no reference of their own.}
\label{fig:recovery}
\end{figure}

\paragraph{Recovery and cooperation gain.}
Recovery is the core capability a detection-only alarm lacks, and it propagates trust
beyond the anchored drones. Figure~\ref{fig:recovery} sweeps the covert ramp over a
low-rate ($0$--$1.0$\,cm/s), long-horizon ($400$\,s) Tier-2 SITL harvest (ArduPilot
EKF/controller-in-the-loop; median over $10$ seeds/rate)---a deliberately slower,
longer regime than the $2$--$20$\,cm/s, $90$\,s harvest used for the onboard monitor
(App.~\ref{app:monitor}). The reported-vs-true position error grows with the attack---reaching
$3.6$\,m at the top swept rate ($1.0$\,cm/s over the $400$\,s horizon)---while anchor-rooted recovery holds every
drone near the anchor-noise-limited floor ($\approx\!0.5$\,m) regardless of rate: anchored drones to
$0.45$\,m and, crucially, the non-anchored drones, which carry no absolute
reference of their own, to $0.56$\,m via range-based trust propagation. (These harvest
flights run longer than the Tier-1 horizon of Table~\ref{tab:params}, so the
absolute drift exceeds $v\,(T-t_s)$ under those defaults; the scientific point is
the \emph{flat} recovery curve, not the absolute magnitude.) Recovery overtakes
the raw report once the drift clears its floor and stays flat thereafter.
Table~\ref{tab:defres} summarizes the remaining defenses. Against an aliased
attack, the bootstrap separates a null from the attack residual on the
rendered-vision anchor ($0.23$ vs.\ $2.22$\,m; single Gazebo run);
the joint estimator, with no clean-epoch label, recovers the onset to a mean
absolute error of $0.4$--$3.4$\,s and the ramp rate to $0.2$--$0.5$\,cm/s, and lifts
change-point detection AUC from $0.00$ (naive) to $1.00$---provided the pre-onset
window is $\gtrsim\!10$\,s; below that the drift/attack collinearity inflates the
ramp error to $\sim\!10$\,cm/s. Under simulator-injected Byzantine anchors (Tier-2,
ArduPilot with a lying second GNSS; $20$ runs/level, bootstrap over independent
runs), robust recovery holds at $25\%$ compromise ($0.31$\,m) and collapses at
$50\%$ ($2.27$\,m), matching the barrier of \S\ref{sec:defense} and the Tier-1
ablation below.

\begin{table}[t]
\caption{Defense results across mechanisms (all simulation); the evaluation
metric differs per row (second column).}
\label{tab:defres}
\small
\begin{tabular}{@{}p{2.3cm}p{2.05cm}p{1.95cm}l@{}}
\toprule
\textbf{Mechanism (setting)} & \textbf{Metric} & \textbf{Result} & \textbf{$n$} \\
\midrule
Bootstrap drift-corr.\ (aliased, Gazebo) & residual, null/attack & $0.23/2.22$\,m & 1 run \\
Joint est.\ (Tier-1) & onset MAE & $0.4$--$3.4$\,s & 40/cond. \\
Joint est.\ (Tier-1) & ramp-rate MAE & $0.2$--$0.5$\,cm/s & 40/cond. \\
Byzantine recovery (SITL) & recovery err., $25/50\%$ & $0.31/2.27$\,m & 20/lvl \\
8-vehicle SITL swarm & recovery / GPS drift & $0.39/10.1$\,m & 20 \\
Rendered-vision multi-SITL & recovery / GPS drift & $0.071/3.2$\,m & 5 \\
Vision tracking (in coverage) & track err.; envelope & $\approx\!6$\,cm; $\approx\!4$\,m & 8/pt \\
\bottomrule
\end{tabular}
\end{table}

\begin{figure*}[t]
\centering
\includegraphics[width=0.85\textwidth]{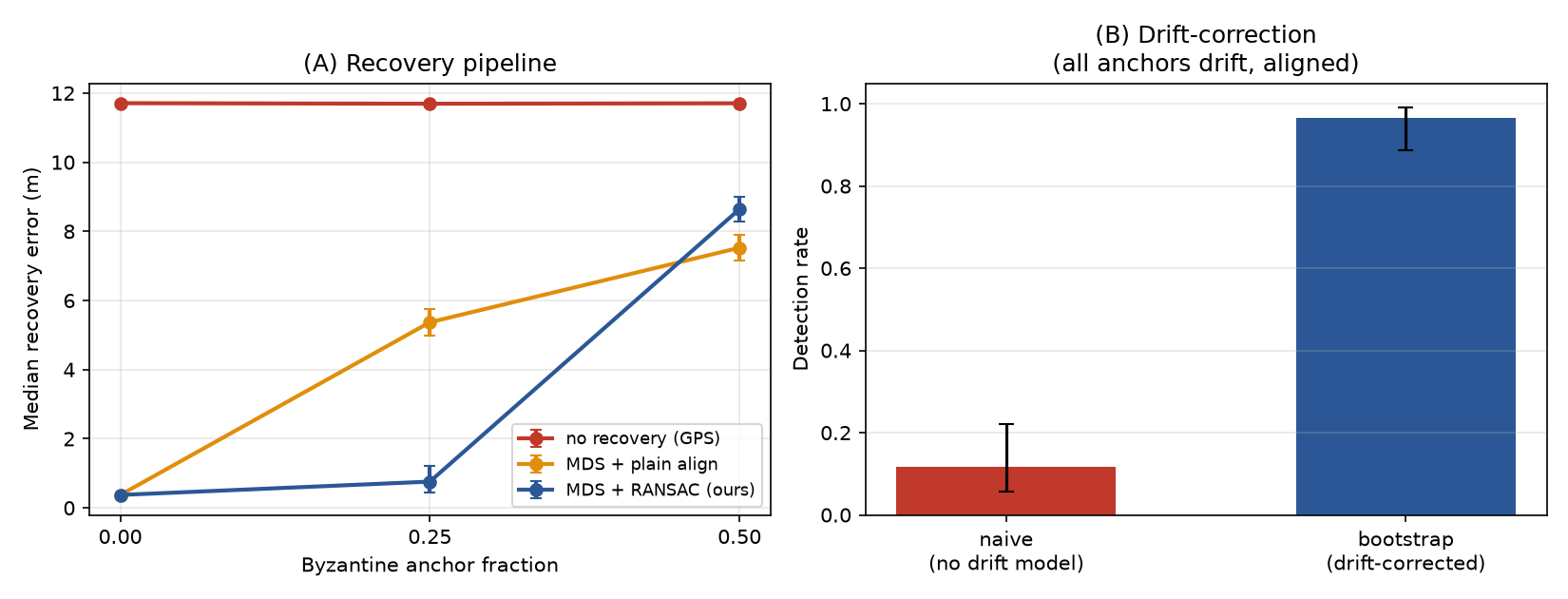}
\caption{Component ablation (Tier-1, $60$ seeds, $95\%$ CIs). (A) MDS$+$align
provides recovery ($11.7\!\to\!0.37$\,m); RANSAC's benefit is specific to a
Byzantine \emph{minority} ($25\%$: $5.4\!\to\!0.76$\,m) and disappears at the
$50\%$ barrier. (B) Bootstrap drift-correction restores detection
($0.12\!\to\!0.97$) when all anchors drift aligned with the attack.}
\label{fig:ablation}
\end{figure*}

\paragraph{Component ablation.}
To separate what each stage buys, we ablate on the same Tier-1 rigid-covert
scenario with $95\%$ CIs over $60$ seeds (Figure~\ref{fig:ablation}). \emph{(A)}
Recovery: with no liars, MDS shape $+$ anchor alignment already cuts the median
error from the $11.7$\,m GPS drift to $0.37$\,m, and RANSAC adds nothing yet. Its
value is specific to a Byzantine \emph{minority}---at $25\%$ Byzantine anchors, plain
least-squares alignment is dragged to $5.4$\,m while RANSAC holds at $0.76$\,m---and
vanishes at $50\%$, where no majority exists and RANSAC is no better than (indeed
slightly worse than) plain alignment: the barrier of \S\ref{sec:defense} made
concrete. \emph{(B)} Drift-correction: when \emph{all} anchors drift aligned with
the attack, the naive median residual is masked (detection $0.12$), while the
clean-epoch bootstrap restores it to $0.97$. Each component's benefit is thus
localized to the regime it targets, not an undifferentiated stack. The
simulator-injected Tier-2 run ($20$ seeds) reproduces the same qualitative boundary as panel (A)---RANSAC
robust recovery $0.31$\,m under $25\%$ Byzantine anchors but $2.27$\,m at $50\%$---so the
component boundary is not a Tier-1 artifact.

\paragraph{Additional threat and sensing conditions.}
Appendix~\ref{app:supp} evaluates the defense under further conditions. As a
by-product, per-drone \emph{attribution} of a
partial (subset) spoof---a distinct, geometry-breaking attack---becomes reliable
above a $\approx\!1.5$\,m offset (App.~Fig.~\ref{fig:attribution}). A single
\emph{heterogeneous} modality resolves the otherwise-ambiguous common-mode
attack-vs-anchor-fault case ($40/40$; Caveat~1, App.\ Fig.~\ref{fig:distinguish}).
Non-anchored recovery stays $\le\!0.5$\,m under the evaluated heavy-tailed
real-UWB errors (IDLab Ghent~\cite{fontaineuwb}; $0.34\!\to\!0.39$\,m, detection
still $1.0$) and sparse range graphs (down to $\approx\!35\%$ density;
App.~Fig.~\ref{fig:sparse}). Against an adaptive attacker, a back-loaded ramp only
delays time-to-detect ($5.7\times$) and the minimax bootstrap cuts undetected drift
from $4.80$ to $0.36$\,m, raising the attacker's required capability to
active-majority anchor compromise or the simultaneous-isomorphic
$\tau\!\to\!0$ aliasing barrier (\S\ref{sec:barrier}). Runtime
(per-frame detection flat in $N$; Byzantine-robust recovery is the scaling cost,
App.~Table~\ref{tab:runtime}) and a distribution-free detection certificate are
also reported there.

\subsection{RQ3: Closed-loop multi-SITL validation}
\begin{figure}[t]
\centering
\includegraphics[width=0.82\columnwidth]{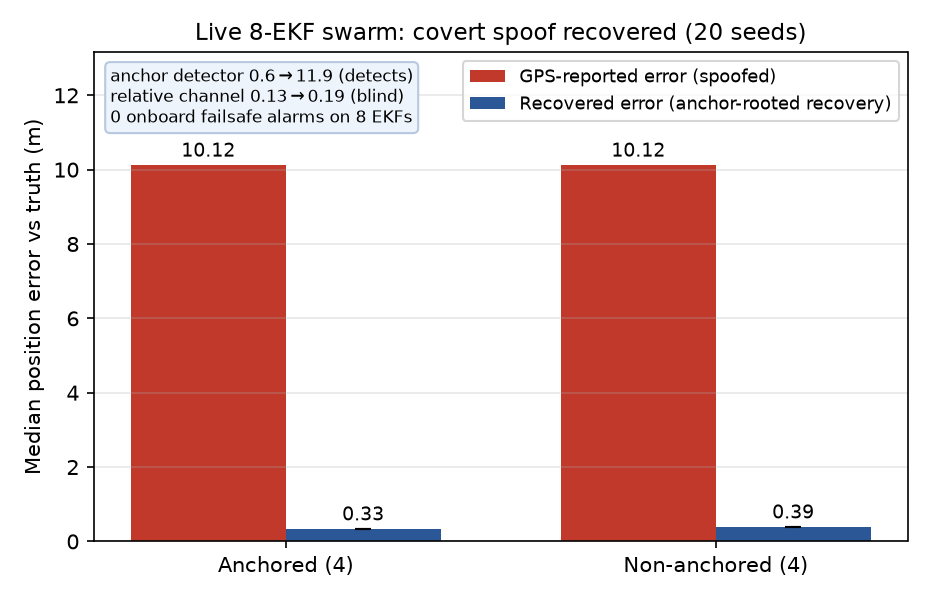}\\[1pt]
{\footnotesize (a) Eight-vehicle whole-swarm recovery ($20$ seeds)}\\[4pt]
\includegraphics[width=0.82\columnwidth]{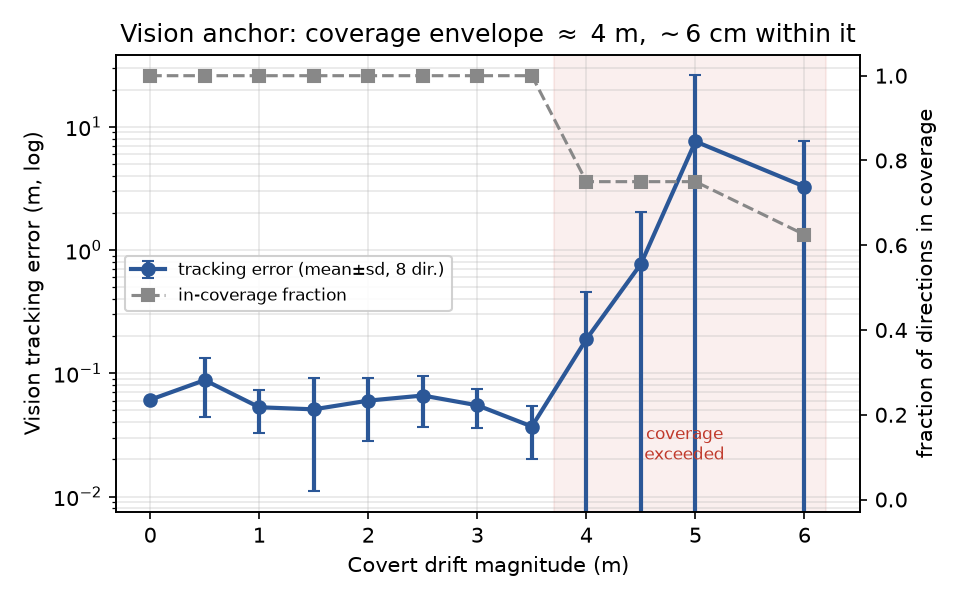}\\[1pt]
{\footnotesize (b) Vision-anchor coverage envelope ($8$ directions/point)}
\caption{Closed-loop multi-SITL validation. (a) Under a common-mode covert spoof,
GNSS drifts by $\approx 10.1$\,m while anchor-rooted recovery reconstructs all
eight positions, including the four non-anchored drones, to a median error of at most $0.39$\,m
($20$ seeds, $4$ anchors). (b) The Gazebo-rendered vision anchor maintains
$\approx\!6$\,cm tracking error within the fully covered region (below the
$\approx\!4$\,m boundary) and degrades sharply beyond it, as landmarks leave the
field of view. The rendered-vision
multi-SITL capstone ($5$ seeds) is evaluated within this measured operating
envelope.}
\label{fig:swarm8}
\end{figure}

\paragraph{Closed-loop single-drone case study.}
As an illustrative closed loop (ArduPilot coupled to Gazebo physics), a covert
GNSS ramp drives one drone $-1.10/-3.38/-5.62$\,m of simulated displacement (slope
$1.01$ vs.\ injection) while GNSS/telemetry reports $0$\,m with no built-in
glitch/failsafe alarm; the
Gazebo vision anchor (HSV$+$\texttt{solvePnP}, an actual image pipeline with
non-Gaussian failure modes, not ``true value $+$ Gaussian noise'') tracks the
drift to $4$--$7$\,cm within its coverage envelope (Figure~\ref{fig:swarm8}b) and
degrades sharply beyond it, and an abrupt spoof removal trips the built-in glitch
gate while the gradual ramp evades it (consistent with Caveat~2;
App.~Fig.~\ref{fig:e58}). The seeded, quantitative version is the eight-vehicle
swarm.

\paragraph{Eight-vehicle SITL swarm.}
Scaling to eight ArduPilot SITL vehicles (the ArduPilot EKF3 $+$ controller;
$20$ seeds, $4$ non-collinear anchors, $\rho{=}0.5$), a common-mode covert ramp
($20$\,cm/s) drifts all eight drones $\approx 10.1$\,m while \emph{zero}
built-in onboard glitch/failsafe alarms fire on any vehicle in any run (the
offline raw-innovation monitor of \S\ref{sec:intro} would fire, but it is not
part of the deployed autopilot). The relative channel stays
blind (baseline score $0.13\!\to\!0.19$), the anchor detector fires
($0.62\!\to\!11.9$), and anchor-rooted recovery reconstructs absolute-position estimates for all
eight drones---the four anchored drones to $0.33$\,m and, crucially, the four
non-anchored drones (with no absolute reference of their own) to $0.39$\,m against
the $10.1$\,m GPS drift (Figure~\ref{fig:swarm8}a). Bootstrapping over the $20$
independent runs (unit $=$ one run, each with its own SITL trajectory and
sensor-noise draw; \S\ref{sec:impl}), the recovery is run-to-run stable---not a single
lucky run.

\paragraph{Coverage envelope and rendered-vision capstone.}
The rendered-vision anchor is accurate only within its measured visual coverage
envelope: Figure~\ref{fig:swarm8}(b) shows $\approx\!6$\,cm tracking error below
the $\approx\!4$\,m boundary, rising sharply beyond it as landmarks leave the
field of view. We
therefore evaluate the multi-SITL capstone only \emph{within} this operating
envelope, rather than assuming an unlimited synthetic anchor. Under that
condition---rendering the SITL swarm's anchor drones through the vision pipeline
over $5$ independent seeds, using the ArduPilot EKFs and Gazebo renders with no
synthetic anchor channel---the vision-pipeline anchors recover their own absolute
positions to $\approx\!5$\,cm (median per seed, sd $1.2$\,cm) and anchor-rooted
recovery reconstructs the four non-anchored drones to $7.1$\,cm (sd $1.2$\,cm)
against a $3.2$\,m GPS drift. (This capstone and the eight-vehicle result of
Figure~\ref{fig:swarm8}(a) are distinct settings---$5$ vs.\ $20$ seeds, $3.2$ vs.\
$10.1$\,m of drift---not a single run.) The four trusted anchors are
\emph{non-collinear} by design: a $3$-anchor set placed on the formation's
collinear first row leaves the MDS reflection unresolved
(\S\ref{sec:arch}) and can flip the non-anchored recovery, whereas $4$
non-collinear anchors ($\rho{=}0.5$) resolve it---the configuration we report.

\section{Related Work}\label{sec:related}

\paragraph{GNSS spoofing attacks.}
Portable civilian spoofers~\cite{humphreys2008}, the requirements for successful
spoofing~\cite{tippenhauer2011}, and field capture-and-control of
UAVs~\cite{kerns2014,noh2019} establish the single-receiver threat; surveys cover
detection broadly~\cite{psiaki2016}. For swarms, GNSS-WASP~\cite{gnsswasp}
demonstrates wide-area formation-preserving spoofing on SDR hardware; the
countermeasures it bypasses---multi-receiver checks with known inter-receiver
distances, and unpredictable-movement/inertial consistency checks---key on the residual an \emph{imperfect} wide-area spoof leaves
across separated receivers, so they do not address the idealized \emph{perfectly
common} translation we study, which leaves no such per-receiver bias. Formal
analyses bound a coordinated single spoofer to $\le 9$ formation-preserving
receivers with bounded residual beyond~\cite{chen2025} and propose
no defense; SwarmFuzz~\cite{swarmfuzz2026} \emph{finds} propagation vulnerabilities
but breaks geometry, the opposite of \rgd.

\paragraph{Single-vehicle sensor attacks, detection, and recovery.}
Sensor attacks span acoustic gyroscope injection~\cite{son2015}, IMU
spoofing~\cite{tu2018}, GPS input spoofing~\cite{davidson2016}, and
Kalman-defeating multi-sensor signal injection~\cite{nashimoto2018}. Control- and
invariant-based detectors~\cite{choi2018,quinonez2020} and their stealthy-attack
limitations~\cite{dash2019,khazraei2024} are well studied, as is attack
\emph{recovery}~\cite{choi2020ssr,dash2021,kong2018,zhang2020,zhang2021} and,
recently, diagnosis-guided recovery from multi-sensor deception in
RVs~\cite{dash2024}. Vision/imagery anchors for single-UAV spoofing detection
include DeepSIM~\cite{xue2020}. All of these are single-vehicle: they neither
model a swarm's gauge freedom nor recover a whole formation's true positions from
a trusted subset.

\paragraph{Cooperative swarm defenses.}
The defenses of Table~\ref{tab:defenses}~\cite{swarmraft,bi2023,bdgd,michieletto2023,orbitguardnet,tristream,meng2023}
key on relative geometry and carry no absolute reference; several explicitly
leave the rigid or recovery case open. Pure multi-receiver relative
verification~\cite{tippenhauer2016} likewise collapses a single-transmitter
attacker to one point and detects the resulting baseline break, but is blind to a
formation-preserving common offset. Byzantine-resilient
consensus~\cite{leblanc2013}, RANSAC~\cite{fischler1981}, MDS, and
CUSUM~\cite{page1954} are standard mechanisms we build on, not contributions.

\paragraph{Detection limits and drift learning.}
CPS attack identification~\cite{pasqualetti2013} and impact
bounds~\cite{murguia2017} give the general theory; Khazraei et
al.~\cite{khazraei2024} prove single-vehicle stealthy-attack existence; and the
closest analog to our law, Baweja~\cite{baweja2026}, proves a timing-domain
slow-ramp impossibility. On the estimator side, blind drift
calibration~\cite{prnet} removes benign drift (the inverse of our goal) and a
single-UAV change-point detector over an internal RL signal~\cite{panda2025} is
the closest adjacency; neither performs joint $(g,v,t_s)$ estimation across a
swarm of simultaneously-drifting absolute anchors.

\paragraph{Delta.} The observability propositions and the mechanisms are prior
art and cited as such. Our original elements are the empirical anchor
failure-mode characterization (\S\ref{sec:anchorfail}), the drift-proportional
detection-limit law in the swarm/anchor/position setting (\S\ref{sec:limit}), the
joint change-point estimator over the aggregated drifting-anchor residual
(\S\ref{sec:est}), and the rendered-vision multi-SITL instantiation
(\S\ref{sec:eval}).

\section{Discussion, Limitations, and Ethics}\label{sec:limits}
\paragraph{Limitations.} All results are in simulation; there is no RF spoofing
hardware and no physical swarm, and the vision anchor and flights are
Gazebo-rendered (the flights use the ArduPilot EKF and controller). Beyond that,
the results rest on assumptions we make explicit. \emph{Modeling:} a $2$D
horizontal plane; recovery assumes an \emph{independent, honest} absolute-anchor
channel and a UWB ranging channel the attacker cannot forge, with sufficiently
accurate time synchronization. \emph{Recovery geometry:} classical MDS needs a
\emph{connected} range graph, at least three \emph{non-collinear} honest anchors
to remove the reflection ambiguity (\S\ref{sec:arch}), and $m\ge\max(2f{+}1,f{+}3)$
under $f$ Byzantine anchors; sparse graphs are handled by geodesic completion,
which introduces distance distortion and fails once the graph disconnects
(\S\ref{sec:eval}). \emph{Attack model:} the detection-limit law is derived for a
constant-rate ramp (nonlinear profiles need a separate analysis); the $N{=}128$
result is defense scalability under an \emph{ideal} distributed-spoofer oracle, not
a validated wide-area attack; and a homogeneous common-mode anchor fault is
attributable only with a heterogeneous modality (Caveat~1). \emph{Estimator scope:}
the joint estimator assumes the aggregated honest-anchor drift is well approximated
by an affine trend and that the attack introduces a single resolvable change point;
heterogeneous nonlinear drifts and multiple change points are not evaluated. \emph{Anchor:} vision
recovery is bounded by landmark coverage and visibility. Two barriers are
fundamental: an attack simultaneous and isomorphic to the drift
($\tau\!\to\!0$), and active compromise of a majority of anchors. The
observability propositions are prior-theory instantiations, not new theorems; our
contributions are the quantitative detector-specific limit, the joint estimator,
and the system evaluation.

\paragraph{Deployment.} The defense needs a trusted absolute-reference modality
on a subset of drones; vision or terrain matching is one realization; fixed beacons
or a separately authenticated navigation source whose signal path lies outside
the attacker's control are others. Coverage (\S\ref{sec:anchorfail};
Fig.~\ref{fig:swarm8}b) and
anchor ratio $\rho$ (\S\ref{sec:limit}) are the key design parameters.
\emph{Runtime:} detection stays real-time at all evaluated sizes, but the current
single-core exhaustive-RANSAC recovery does not sustain the $10$\,Hz update rate at
$N{=}64$ ($139$\,ms/frame, App.~Table~\ref{tab:runtime}); Byzantine-robust recovery
at that scale needs optimization, parallelization, or a lower update frequency,
and the $N{=}128$ result demonstrates \emph{detection} scalability, not Byzantine-robust
recovery throughput.

\paragraph{Ethical considerations.} This work is defense-oriented and introduces
no offensive capability beyond publicly established
results~\cite{gnsswasp,chen2025,khazraei2024}. All experiments are simulated: no
real GNSS signals are transmitted and no physical aircraft are flown, so there is
no spectrum interference or flight-safety risk. We see no human-subjects concern.

\section{Conclusion}
Rigid common-mode GNSS spoofing exposes a structural blind spot in the
\emph{relative-geometry} channel used by the cooperative defenses we examine
(Prop.~1). An independent absolute reference breaks this gauge freedom, and our
anchor-rooted pipeline uses a trusted subset to recover the positions of
the entire swarm, including non-anchored drones. We quantify the detector's
drift-dependent floor and use temporal estimation to separate anchor drift from a
later change-point attack. In eight-vehicle ArduPilot SITL, the method reduces
approximately $10.1$\,m of GNSS error to $0.39$\,m for non-anchored drones, and
the rendered-vision multi-SITL experiment achieves a median recovery error of
$7.1$\,cm under $3.2$\,m of drift. The remaining limits---non-collinear anchor geometry, anchor
coverage, $\tau\!\to\!0$ aliasing, and majority anchor compromise---are explicit
rather than hidden assumptions.

\bibliographystyle{ACM-Reference-Format}
\bibliography{refs}

\appendix
\section{Open Science}\label{app:repro}
In keeping with the ACM open-science policy, all code, configurations, summary
data, and figure-generation scripts are released; the artifact is publicly available at \url{https://github.com/rokyp1278/RigidShift}. Tier-1 results are seed-fixed and byte-identical on a fixed dependency set, and
reproducible to floating-point tolerance ($\le\!10^{-13}$ relative) across
NumPy/BLAS versions. Configurations, summary
metrics, and figures are versioned and regenerated from committed result files by
a single script; the ArduPilot/Gazebo recipe, the multi-vehicle harvest harness,
and the closed-loop and multi-SITL walkthroughs are provided with the artifact.
Detector code is shared across tiers via a common schema so synthetic and
simulator data exercise identical logic.

\section{Offline EKF-Innovation Monitor}\label{app:monitor}
The \S\ref{sec:intro} claim that a non-geometric onboard channel can catch the
covert ramp is supported by a purpose-built \emph{offline} monitor over the
per-drone EKF position-innovation sequence---an onboard-sensor detector in the spirit of~\cite{b1onboard}, distinct from the
autopilot's built-in glitch/failsafe gate. The statistic is the gated peak of
the smoothed absolute innovation; the threshold is calibrated at the $5\%$
false-alarm quantile of the no-attack runs. On the Tier-2 ArduPilot SITL harvest
it detects the common-mode covert ramp on $20/20$ attack runs at every tested rate
($2$, $5$, $10$, $20$\,cm/s; Clopper--Pearson $95\%$ CI $[0.84,1.0]$). The
corresponding held-out (leave-one-out) false alarm was $1/10$ no-attack runs
($10\%$, roughly double the $5\%$ target)---a small-sample estimate with a wide
finite-sample interval, and precisely the kind of unverifiable false-alarm burden
a lone drone cannot escape. This is in any case a \emph{bare alarm}: it recovers
no true positions, which is the gap our
anchor-rooted recovery fills. (A finer rate sweep down to $0.25$\,cm/s shows the
monitor sits near its floor---$1/10$ at $0.25$\,cm/s, $1/10$ at
$0.5$\,cm/s, $10/10$ at $1.0$\,cm/s---whereas the anchor detector stays at $1.0$
throughout.)

\section{Supporting Figures and Analyses}\label{app:supp}
For readability the main text keeps one representative number per result; the detailed curves, sweeps, and secondary experiments live here. All are produced using the same code and configurations as the main-text results and do not change any of the conclusions. Table~\ref{tab:secondary} collects the secondary numbers the main text quotes once, each tagged with the experiment that produces it (standalone plots are in the artifact).

\begin{table}[H]
\caption{Secondary results quoted once in the main text, each with the experiment
identifier that produces it in the artifact.}
\label{tab:secondary}
\small
\begin{tabular}{@{}p{3.85cm}p{2.75cm}l@{}}
\toprule
\textbf{Result} & \textbf{Value} & \textbf{Exp.} \\
\midrule
Heavy-tailed real-UWB recovery & $0.34\!\to\!0.39$\,m, det.\ $1.0$ & e65 \\
Back-loaded ramp time-to-detect & $5.9\!\to\!33.4$\,s ($5.7\times$) & e32 \\
Minimax bootstrap undetected drift & $4.80\!\to\!0.36$\,m & e47 \\
Distribution-free certified floor $v_{\text{cert}}$ & naive $4.0$--$10.0$; bootstrap $5.0$\,cm/s & e46 \\
Realistic GNSS-WASP residual, rel.-channel det. & $0.07$ ($1\times$) $\to$ $1.0$ ($500\times$) & e9 \\
\bottomrule
\end{tabular}
\end{table}

\paragraph{Distribution-free detection certificate.} For a calibrated anchor
detector whose per-run score has mean $\mu(v)$ and variance $\sigma^2$ at
threshold $\mathrm{thr}$, a one-sided Cantelli (Chebyshev) inequality gives the
distribution-free guarantee $\Pr[\text{detect}\mid v]\ge a^2/(\sigma^2+a^2)$ with
$a=\mu(v)-\mathrm{thr}$. Using conservative moment estimates (a mean lower and a
variance upper confidence bound) yields a certified minimum detectable rate
$v_{\text{cert}}$ (Table~\ref{tab:secondary}); on disjoint held-out runs the
certified bound was never violated.

\begin{figure}[H]
\centering
\includegraphics[width=\columnwidth]{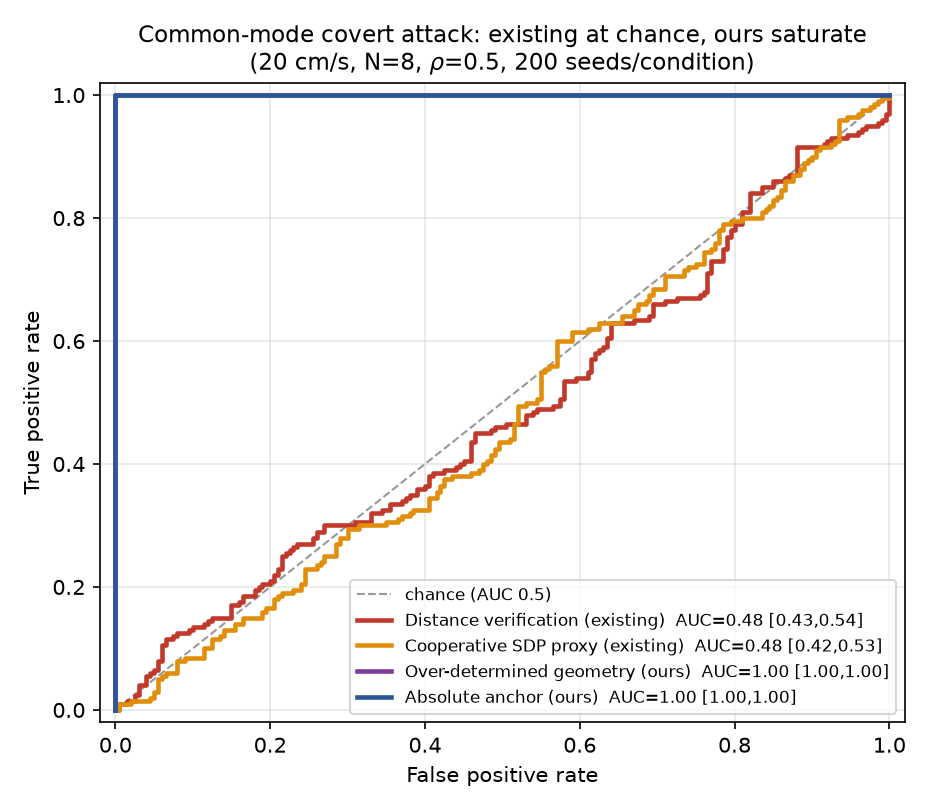}
\caption{Combined ROC under the common-mode covert attack ($20$\,cm/s, $N{=}8$,
$\rho{=}0.5$, $200$ seeds/condition, bootstrap $95\%$ CIs). Both existing
relative detectors sit at chance (CI includes $0.5$); our over-determined
geometry and absolute anchor reach AUC $1.0$ above $v^{\ast}$.}
\label{fig:auc}
\end{figure}

\begin{figure}[H]
\centering
\includegraphics[width=\columnwidth]{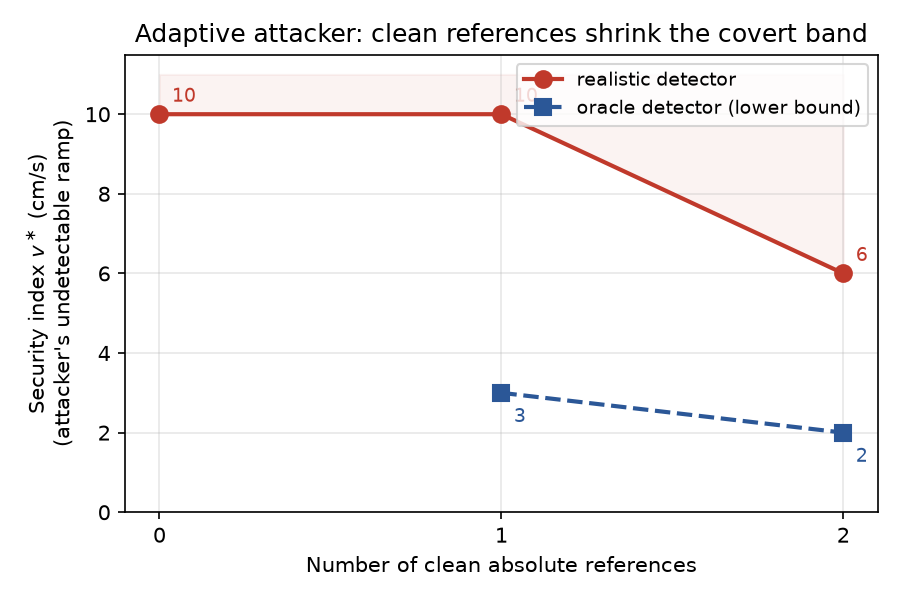}
\caption{Security index: the attacker's undetectable ramp rate $v^{\ast}$ shrinks
as the defender adds clean absolute references ($10\!\to\!6$\,cm/s realistic, down to
$2$\,cm/s for an oracle); the oracle is a lower bound.}
\label{fig:secidx}
\end{figure}

\begin{figure}[H]
\centering
\includegraphics[width=\columnwidth]{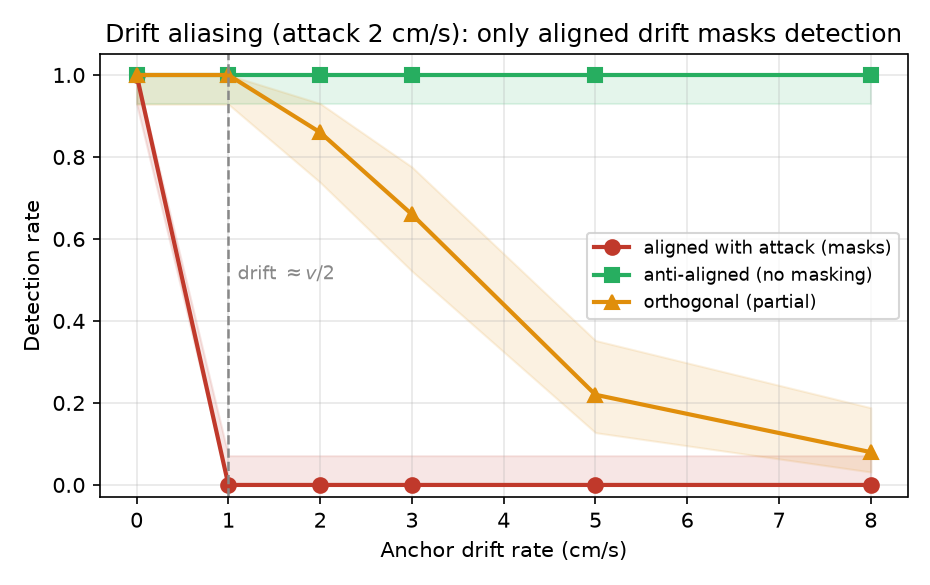}
\caption{Drift aliasing (attack $2$\,cm/s). Only anchor drift \emph{aligned} with
the attack ($\approx v/2$) masks a memoryless detector; anti-aligned drift does
not, and orthogonal drift only partially.}
\label{fig:aliasing}
\end{figure}

\begin{table}[H]
\caption{Per-frame runtime (ms, single core). Detection is flat in $N$; RANSAC
robust recovery is the scaling cost.}
\label{tab:runtime}
\small
\begin{tabular}{@{}lccccc@{}}
\toprule
\textbf{Stage} & $N{=}4$ & $8$ & $16$ & $32$ & $64$ \\
\midrule
Detection (anchor score)   & 0.013 & 0.016 & 0.013 & 0.019 & 0.018 \\
Recovery (MDS, non-robust) & 0.003 & 0.043 & 0.063 & 0.47  & 0.94  \\
Recovery (RANSAC robust)   & 0.002 & 0.15  & 1.61  & 15.3  & 139   \\
\bottomrule
\end{tabular}
\end{table}

\begin{figure}[H]
\centering
\includegraphics[width=\columnwidth]{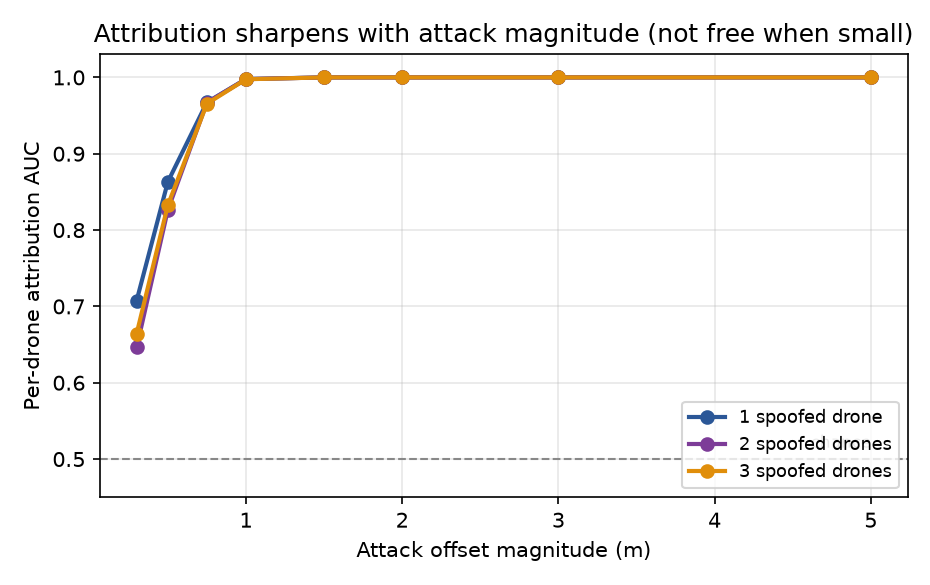}
\caption{Per-drone attribution AUC vs.\ attack magnitude, for $1$--$3$
simultaneously spoofed drones (Tier-1). Attribution saturates at $1.0$ only past
an offset of $\approx 1.5$\,m and degrades toward chance for small offsets or more
spoofed drones.}
\label{fig:attribution}
\end{figure}

\begin{figure*}[t]
\centering
\includegraphics[width=0.92\textwidth]{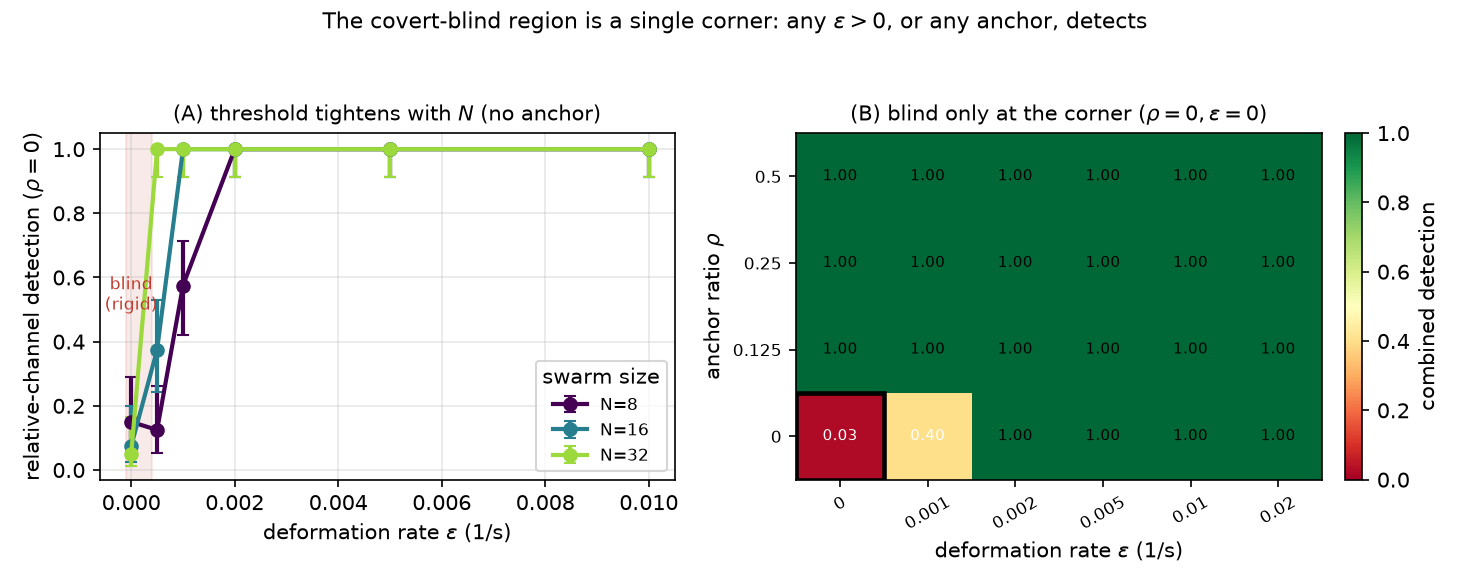}
\caption{The covert-blind region is a single corner. \emph{(A)} With no anchor
($\rho{=}0$), any deformation $\varepsilon\ge0.002$ self-reveals through the
relative channel, and the threshold \emph{tightens} as the swarm grows
($\varepsilon\ge0.002$ at $N{=}8$, $\ge0.001$ at $N{=}16$, $\ge0.0005$ at $N{=}32$;
$40$ seeds, Wilson CIs). \emph{(B)} Over the $(\rho,\varepsilon)$ grid, detection is
blind only at the exact corner ($\rho{=}0,\varepsilon{=}0$); a single anchor
($\rho{=}0.125$) closes it across all $\varepsilon$.}
\label{fig:frontier}
\end{figure*}

\begin{figure}[H]
\centering
\includegraphics[width=\columnwidth]{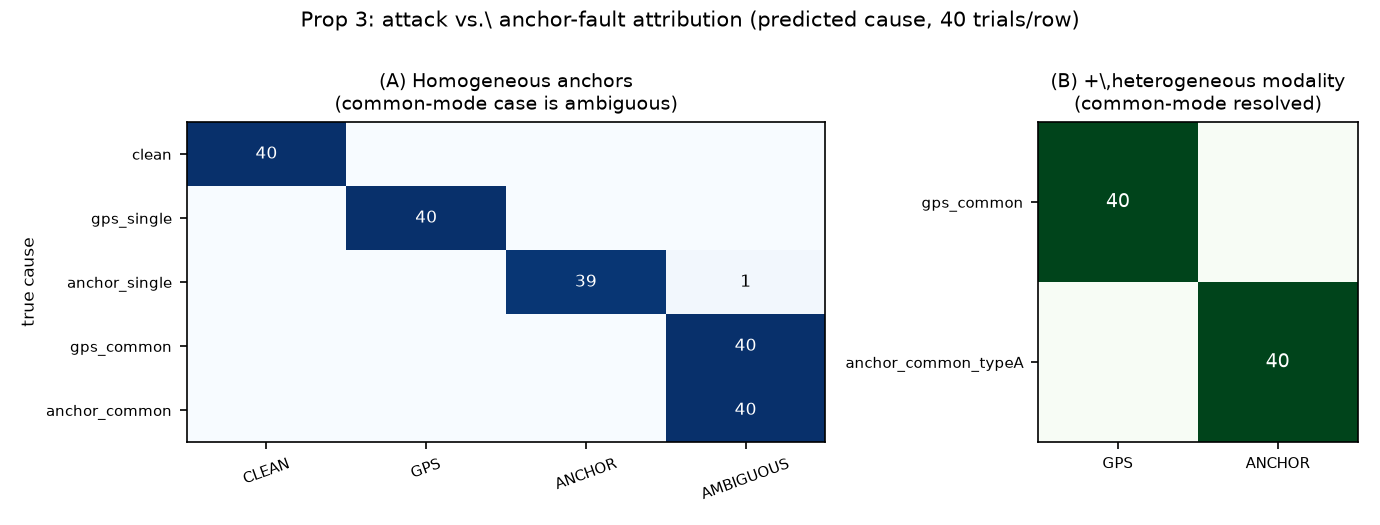}
\caption{Attack vs.\ anchor-fault attribution (Caveat~1, predicted cause, $40$
trials/row). (A) Homogeneous anchors leave a common-mode case ambiguous; (B) one
heterogeneous modality attributes the common-mode GPS attack and anchor fault
correctly.}
\label{fig:distinguish}
\end{figure}

\begin{figure}[H]
\centering
\includegraphics[width=\columnwidth]{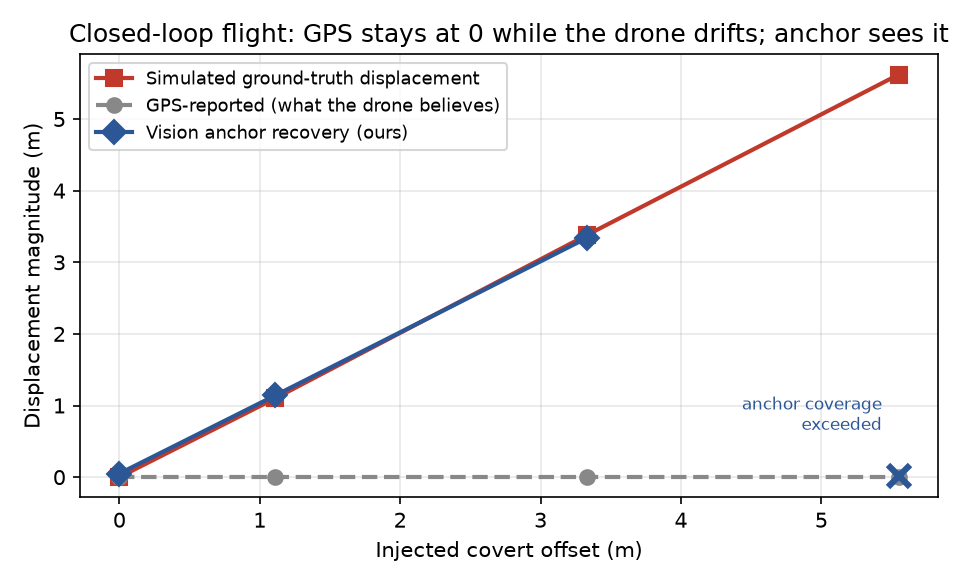}
\caption{Closed-loop SITL flight (ArduPilot\,$+$\,Gazebo): GNSS-reported
displacement stays at $0$ while the drone's simulated position drifts; the vision
anchor tracks the true drift until its coverage is exceeded.}
\label{fig:e58}
\end{figure}

\begin{figure*}[t]
\centering
\includegraphics[width=0.65\textwidth]{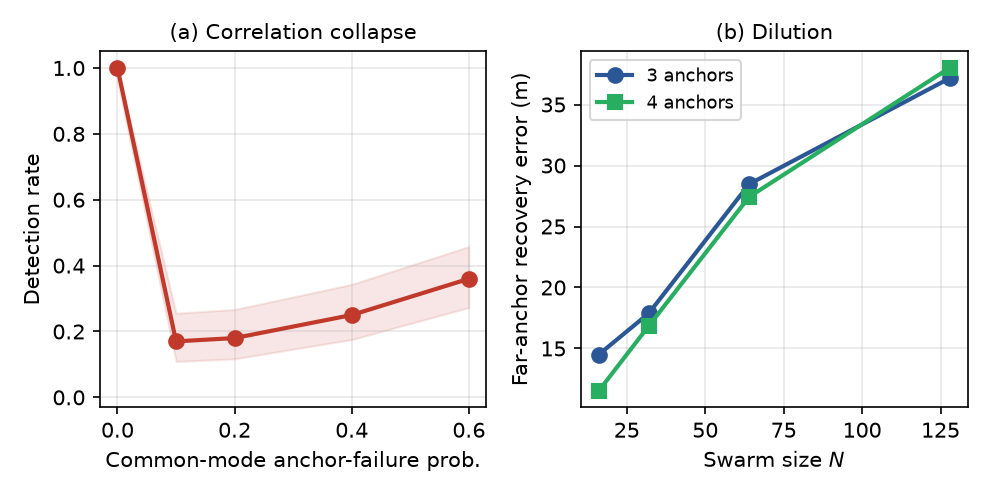}
\caption{Two anchor-failure modes. \emph{(a)} Correlated (common-mode) anchor
failure collapses detection from $1.0$ to $0.17$--$0.36$ as anchors increasingly fail together ($0.17$ at $10\%$ co-failure),
far sharper than independent dropout. \emph{(b)} With a fixed anchor count,
far-from-anchor recovery error grows with swarm size (dilution); adding a fourth
anchor barely helps, so the anchor \emph{ratio} governs recovery.}
\label{fig:anchorfail}
\end{figure*}

\begin{figure*}[t]
\centering
\includegraphics[width=0.85\textwidth]{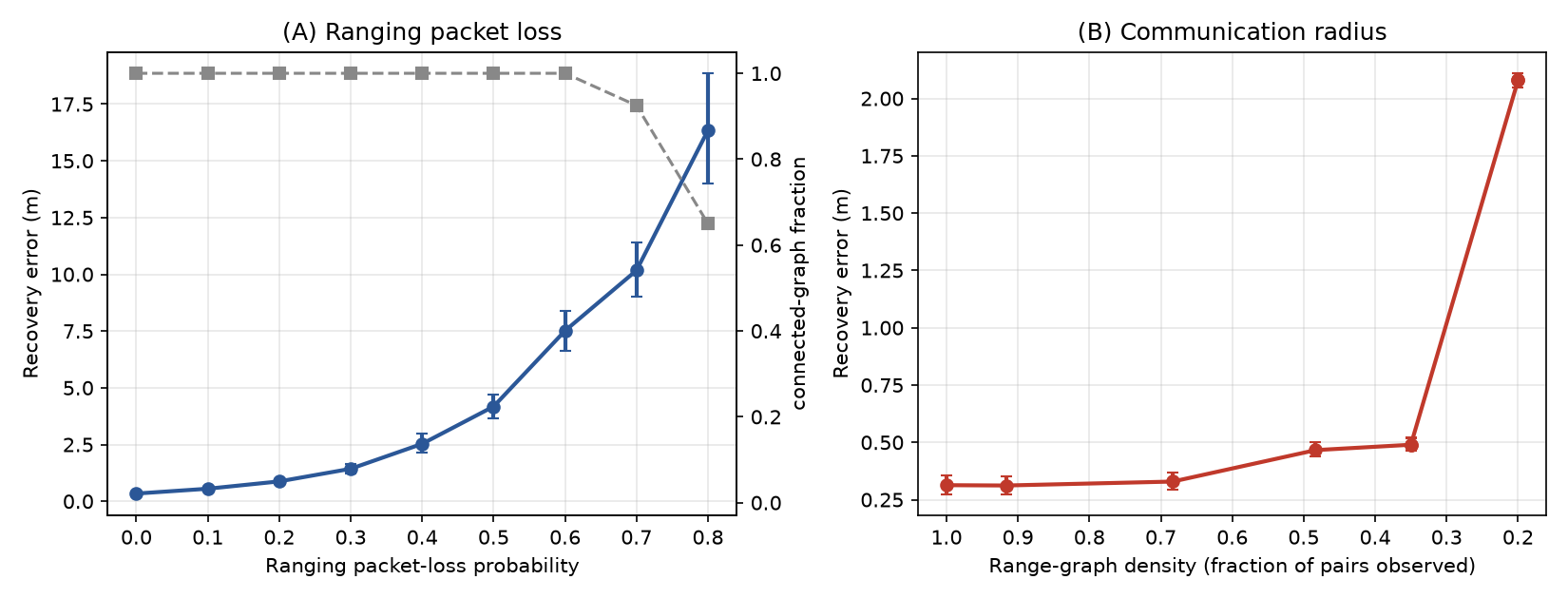}
\caption{Range-graph sparsity ($N{=}16$, geodesic completion, $95\%$ CIs). (A)
Random ranging packet loss degrades recovery and eventually disconnects the graph.
(B) Under a communication radius, recovery holds $\le0.5$\,m down to
$\approx\!35\%$ range-graph density. All-pairs ranging is not required.}
\label{fig:sparse}
\end{figure*}

\end{document}